%% file: ms.tex
 \documentclass[final,authoryear,5p,times]{elsarticle}

\usepackage{graphicx}

\usepackage{amssymb}
\usepackage{amsmath}
\usepackage{aas_macros}
\usepackage{hyperref}
\usepackage{textcomp}
\usepackage{IEEEtrantools}

\input{local-commands}

\journal{Astronomy and Computing}

\begin{document}

\begin{frontmatter}



\title{Rabacus: A Python Package for Analytic Cosmological Radiative Transfer Calculations}


\author[label1]{Gabriel Altay}
\author[label1]{John H. Wise}

\address[label1]{Center for Relativistic Astrophysics, Georgia Institute of Technology, 837 State Street, Atlanta, GA, 30332, USA}

\begin{abstract}
We describe \rabacus, a Python package for calculating the transfer of
hydrogen ionizing radiation in simplified geometries relevant to
astronomy and cosmology.  We present example solutions for three
specific cases: 1) a semi-infinite slab gas distribution in a
homogeneous isotropic background, 2) a spherically symmetric gas
distribution with a point source at the center, and 3)  a spherically
symmetric gas distribution in a homogeneous isotropic background.
All problems can accommodate arbitrary spectra and density profiles as
input.  The solutions include a treatment of both hydrogen and helium,
a self-consistent calculation of equilibrium temperatures, and the
transfer of recombination radiation.  The core routines are written in
Fortran 90 and then wrapped in Python leading to execution speeds
thousands of times faster than equivalent routines written in pure
Python. In addition, all variables have associated units for ease of
analysis.  The software is part of the Python Package Index and the
source code is available on Bitbucket at
\url{https://bitbucket.org/galtay/rabacus}.  In addition, installation
instructions and a detailed users guide are available at 
\url{http://pythonhosted.org//rabacus}.

\end{abstract}


\end{frontmatter}


\section{Introduction}
\label{sec.intro}

The transport of radiation is central to model building in astronomy and cosmology, yet few closed form solutions exist even in highly symmetric situations.   In this work we describe \rabacus, a Python package for calculating equilibrium ionization and temperature states for well posed radiative transfer problems in planar or spherical geometries. 

There is a long history of similar numerical approaches in the field of photon dominated regions or photo-dissociation regions \citep[PDRs, see][for reviews]{Hollenbach_99,Rollig_07}.  
PDRs are usually defined as regions of the interstellar medium in which far ultraviolet photons with energies between 6 and 13.6 eV dominate the thermal and chemical dynamics.  Typical densities and temperatures are $\nH \gtrsim 10$~cm$^{-3}$ and $T \lesssim 5000$~K.  As the main goal of PDR codes is the calculation of emission spectra, they typically include large chemical networks and take account of many line excitation processes.  

The goal of \rabacus\ is to provide a similar tool that is optimized
for the transfer of hydrogen ionizing radiation through cosmological
scale systems.  Common applications include quasar absorption systems
\citep{Wolfe_05,Meiksin_09}, the evaporation of mini-halos during
reionization \citep{Shapiro_04}, and Str\"{o}mgren spheres
\citep{Stromgren_39}.  Typical densities and temperatures in these
regions are $\nH \lesssim 10^{-1}$~cm$^{-3}$ and $T \gtrsim 8000$~K.
This allows us to consider a compact chemical network consisting of
hydrogen and helium and limits the list of physical processes that
dominate the dynamics to photo-ionization, photo-heating, collisional
ionization, collisional excitation, recombination, free-free
radiation, Compton scattering, and Hubble expansion.  

The density and temperature regime treated by \rabacus\ essentially
  excludes the cold inter-stellar medium phase of gas where molecules
  and dust begin to play an important role in the thermal and
  chemical state of hydrogen and helium.  However, cooling from metal line
  emission can play an important role in diffuse enriched gas
  \citep[e.g.][]{Wiersma_09}.  Metal line cooling is not treated by
  \rabacus.  However, its effect on hydrogen and helium can be
  approximated by first calculating the equilibrium temperature of a
  system without metals and then running iso-thermal models with lower
  temperatures.

Photo-ionization and heating rates are calculated in \rabacus\ using a multi frequency reverse ray tracing technique \citep{Altay_13_urch}.  In cases of planar geometry this is simply a matter of evaluating exponential integrals, but in the case of spherical geometries embedded in isotropic radiation backgrounds, ray casting is used. This allows us to use the exact path length through each shell in our radiative transfer solutions. 

The development of \rabacus\ has several purposes.  One is the
  treatment of geometries and radiation fields that are not included
  in many popular codes in use today (for example plane parallel
  radiation incident from both sides of a slab or isotropic
  backgrounds incident on spherical geometries).  In addition we
  wanted to use an object oriented approach coupled with a modern
  language with a large user base and fully developed graphical
  libraries. 
In addition, we wanted to make the isolation of physical processes convenient.  We give special attention to  physical processes that are most often treated in an approximate way in cosmological hydrodynamic simulations: 1) the spectral shape of the ionizing background, 2) the effect of helium on spectral hardening and hydrogen ionization profiles, 3) the effect of equilibrium temperatures vs. fixed temperatures, and 4) the effect of using the on-the-spot approximation vs. calculating the transfer of recombination radiation \citep[e.g.][]{Altay_11, McQuinn_11, Friedrich_12, Rahmati_13, Raicevic_14}.   

We developed the Python package \rabacus\ with the goal of making these solutions both available and readily usable by the astrophysics community at large.  To this end, our project is free, open source, and makes use of a high level programming language with a large user base and well supported graphics packages.  \rabacus\ is available on the Python Package Index%
\footnote{\url{https://pypi.python.org/pypi/rabacus}}; 
the source code is available on Bitbucket\footnote{\url{https://bitbucket.org/galtay/rabacus}},
and a users guide is available through pythonhosted.org%
\footnote{\url{http://pythonhosted.org/rabacus}}.  The software is released under the terms of the FreeBSD license%
\footnote{\url{http://opensource.org/licenses/BSD-2-Clause}}. 
While Python has many benefits, the fact that it is an interpreted
language can make computationally intensive tasks slow compared to
compiled codes.  In order to circumvent this problem we implemented
the core functionality of \rabacus\ in Fortran 90 and wrapped the
resulting code in Python using the Numpy package f2py.  In addition,
all variables in \rabacus\ have units through use of the Quantities
Python
package \footnote{\url{https://github.com/python-quantities/python-quantities}}.
This package will automatically be downloaded during installation.
Comprehensive examples in the users guide cover the use of variables
with units.
The rest of this paper is organized as follows.  In \S 2 we present our notation and review some basic physics. In \S 3 we discuss classes that represent sources of radiation. In \S 4 we describe the single zone solvers supplied by \rabacus. In \S5-8 we discuss the geometric solvers, and in \S9 we conclude.

\section{Basic Physics and Notation}
\label{sec.review}

In this section we review the physics governing the temperature and ionization state of cosmological gas. We will consider the number density of six species and indicate ionization states using Roman numerals  \{ $\nHI$, $\nHII$, $\nHeI$, $\nHeII$, $\nHeIII$, $\nel$ \}.  The free electron number density is related to the other species as,
\begin{equation}
\nel = \nHII + \nHeII + 2 \nHeIII 
\end{equation}
and the mass fraction of helium is given as, 
\begin{equation}
Y \equiv \frac{4 \nHe}{\nH + 4 \nHe}, \quad
\frac{\nHe}{\nH} = \frac{Y}{4(1-Y)}
\label{eqn.Y}
\end{equation}
In what follows, we make use of the following notation,
\begin{eqnarray}
\xHI \equiv \frac{\nHI}{\nH}, \quad 
\xHII \equiv \frac{\nHII}{\nH}, \quad
\xe \equiv \frac{\nel}{\nelmax}, \quad 
\nonumber \\
\xHeI \equiv \frac{\nHeI}{\nHe}, \quad 
\xHeII \equiv \frac{\nHeII}{\nHe}, \quad 
\xHeIII \equiv \frac{\nHeIII}{\nHe}
\label{eqn.xnotation}
\end{eqnarray}
where the electron fraction, $\xe$, is the ratio of the number density of electrons and the maximum possible number density of electrons,
\begin{eqnarray}
\nel &=& \xHII \nH + (\xHeII + 2 \xHeIII) \nHe
\label{eqn.ne}
\nonumber \\
&=& \left[ \xHII + \frac{Y(\xHeII + 2 \xHeIII)}{4(1-Y)} \right] \nH
\end{eqnarray}
\begin{eqnarray}
\nelmax &=& \nH + 2 \nHe = \left[ 1 + \frac{Y}{2(1-Y)} \right] \nH.
\label{eqn.nelmax}
\end{eqnarray}

\subsection{Ionization Evolution} 

Considering the processes of photo-ionization, collisional ionization, and recombination, the coupled evolution equations for the ionization fractions take the following form, 
\begin{eqnarray}
\label{eqn.dxHdt}
\frac{d\xHI}{dt} &=& -(\GHI + \CHI \nel) \xHI + 
\RHII  \nel \xHII 
\nonumber \\
\frac{d\xHII}{dt} &=& - \frac{d\xHI}{dt}
\end{eqnarray}

\begin{eqnarray}
\label{eqn.dxHedt}
\frac{d\xHeI}{dt} &=& -(\GHeI + \CHeI \nel) \xHeI + 
\RHeII  \nel \xHeII 
\nonumber \\
\frac{d\xHeII}{dt} &=& - \frac{d\xHeI}{dt} - \frac{d\xHeIII}{dt}
\nonumber \\
\frac{d\xHeIII}{dt} &=& (\GHeII + \CHeII \nel) \xHeII - 
\RHeIII  \nel \xHeIII  
\end{eqnarray} 

where the $\Gamma$'s represent photo-ionization rates, the $C$'s
collisional ionization rates, and the $R$'s recombination rates. 
Collisional ionization and recombination rates have a non-trivial
dependence on temperature and \rabacus\ provides a simple interface to
load the rate fits presented in \citet{Hui_97}.  Photo-ionization
rates depend on the spectrum of ionizing radiation.  

\begin{eqnarray}
\label{eq.pi_rates}
\GHI &=& c \int_{\nuHIth}^{\infty} \frac{\unu \sigmaHI}{h \nu} 
\exp( -\taunu ) d\nu  
\nonumber \\
\GHeI &=& c \int_{\nuHeIth}^{\infty} \frac{\unu \sigmaHeI}{h \nu} 
\exp( -\taunu ) d\nu 
\nonumber \\
\GHeII &=& c \int_{\nuHeIIth}^{\infty} \frac{\unu \sigmaHeII}{h \nu} 
\exp( -\taunu ) d\nu 
\nonumber \\
\taunu &=& \NHI \sigmaHI + \NHeI \sigmaHeI + \NHeII \sigmaHeII
\end{eqnarray}
where $\unu$ is the energy density per unit frequency in radiation,
$\taunu$ is the frequency dependent optical depth in all species, the
$N_{\rm _X}$ are column densities, the $\sigma_{\rm _X}$ are frequency
dependent photo-ionization cross sections, 
and the $\nu_{\rm _X}^{\rm th}$ are ionization thresholds. Note that
$\unu$ represents an optically thin quantity and $\exp( -\taunu )$
accounts for absorbing material between the source of radiation and the
point at which the photo-ionization rate is being calculated. 

\rabacus\ provides tools for the creation of spectra and access to the cross section fits of \citet{Verner_96}.  The variable $\unu$ serves to characterize the radiation field in a geometry independent way.  It is related to more geometry specific characterizations as follows, 
\begin{equation}
  c \unu = 4 \pi \Jnu = F_{\nu}  = \frac{L_{\nu}}{4 \pi r^2} 
\end{equation}
where $\Jnu$ is the angle averaged specific intensity, $F_{\nu}$ is the energy flux per unit frequency of a plane parallel source, $L_{\nu}$ is the energy luminosity per unit frequency of a point source and $r$ is the distance from the point source.  
It is sometimes convenient to group terms using the following notation, 
\begin{eqnarray}
\mIHI = \GHI + \CHI \nel, && \quad \mRHII = \RHII \nel
\nonumber \\
\mIHeI = \GHeI + \CHeI \nel, && \quad \mRHeII = \RHeII \nel
\nonumber \\
\mIHeII = \GHeII + \CHeII \nel, && \quad \mRHeIII = \RHeIII \nel.
\end{eqnarray}
However, it is important to keep in mind that $\nel$ depends on the ionization 
fractions.  Using this notation we can express these equations in matrix form,

\begin{eqnarray}
\dot{\myvec{x}}_{\rm H} &=& \myvec{M}_{\rm H} \myvec{x}_{\rm H}
\nonumber \\
\myvec{M}_{\rm H} &=& 
\left(
\begin{array}{cc}
 - \mIHI &  \mRHII \\
   \mIHI & -\mRHII
\end{array} \right) 
\end{eqnarray}

\begin{eqnarray}
\dot{\myvec{x}}_{\rm He} &=& \myvec{M}_{\rm He} \myvec{x}_{\rm He} 
\nonumber \\
\myvec{M}_{\rm He} &=& 
\left( 
\begin{array}{ccc} 
 - \mIHeI &  \mRHeII & 0 \\
   \mIHeI & -( \mIHeII + \mRHeII ) & \mRHeIII \\
  0                   & \mIHeII  &  -\mRHeIII
\end{array} \right)
\end{eqnarray}

\subsection{Temperature Evolution} 

The evolution of gas internal energy $u$ is governed by the following equation, 
\begin{eqnarray}
u &=& \frac{3}{2} k_{\rm b} T (\nH + \nHe + \nel)
\\
\label{eqn.dTdt}
\frac{du}{dt} &=& \mathcal{H} - \Lambda_c 
- 3 H k_{\rm b} T (\nH + \nHe + \nel)
\end{eqnarray} 
The first term, $\mathcal{H}$, accounts for photo-heating, the second term, $\Lambda_c$, accounts for radiative cooling, and the final term is due to adiabatic cooling from Hubble expansion.  The first two are always included in the \rabacus\ solvers, but the last is optional.  The photo-heating term 
depends on the spectrum of the radiation field. 
\begin{eqnarray}
\label{eq.ph_rates}
\epHI &=& c \int_{\nuHIth}^{\infty}  \frac{\unu \sigmaHI}{h \nu} 
(h\nu-h\nuHIth) \exp( -\taunu ) d\nu 
\nonumber \\
\epHeI &=& c \int_{\nuHeIth}^{\infty}  \frac{\unu \sigmaHeI}{h \nu}
(h\nu-h\nuHeIth) \exp( -\taunu ) d\nu 
\nonumber \\
\epHeII &=& c \int_{\nuHeIIth}^{\infty}  \frac{\unu \sigmaHeII}{h \nu}
(h\nu-h\nuHeIIth) \exp( -\taunu ) d\nu 
\nonumber \\
\mathcal{H} &=& \epHI \nHI + \epHeI \nHeI + \epHeII \nHeII
\end{eqnarray}
We consider the following radiative cooling processes, 

\begin{itemize}
\item Collisional Ionization Cooling (cic)
\item Collisional Excitation Cooling (cec)
\item Dielectronic Recombination Cooling (di)
\item Radiative Recombination Cooling (re)
\item Bremsstrahlung Cooling (br)
\item Compton Heating/Cooling (cp)
\end{itemize} 
Their contribution to the cooling function is calculated as follows, 

\begin{eqnarray}
\frac{\Lambda_c}{\nel} &=& 
\Lambda^{\rm cec}_{\rm HI} \nHI + 
\Lambda^{\rm cec}_{\rm HeII} \nHeII +  
\Lambda^{\rm di}_{\rm HeII} \nHeII +  
\nonumber \\ 
\frac{}{} && 
\Lambda^{\rm cic}_{\rm HI} \nHI + 
\Lambda^{\rm cic}_{\rm HeI} \nHeI +  
\Lambda^{\rm cic}_{\rm HeII} \nHeII +  
\nonumber \\ 
\frac{}{} && 
\Lambda^{\rm re}_{\rm HII} \nHII + 
\Lambda^{\rm re}_{\rm HeII} \nHeII +  
\Lambda^{\rm re}_{\rm HeIII} \nHeIII +  
\nonumber \\ 
\frac{}{} && 
\Lambda^{\rm br} g_{\rm ff} T^{1/2} ( \nHII + \nHeII + 4 \nHeIII) + 
\nonumber \\ 
\frac{}{} && 
\Lambda^{\rm cp} (1+z)^4 (T - T_{\rm cmb})
\end{eqnarray}
The first nine terms are products of temperature dependent rates and ion
number densities.  The next term accounts for bremsstrahlung cooling and involves a Gaunt factor which we set to $g_{\rm ff} = 1.1 + 0.34 \exp[ -(5.5 - \log_{10} T)^{2/3} ]$.  The final term accounts for Compton cooling.  The fits
used in \rabacus\ for these rates are from \citet{Hui_97}.

\section{Source Classes}
\label{sec.src}

There are three source classes available in \rabacus\, namely, {\tt PointSource}, {\tt PlaneSource}, and {\tt BackgroundSource}.  The sources are grouped by geometry due to the different kinds of integrals that need to be computed in order to calculate quantities like photo-ionization and photo-heating rates.   All classes presented here take a common set of arguments when instantiated.  The first two, {\tt q\_min} and {\tt q\_max} are floats equal to the minimum and maximum photon energies considered divided by Rydbergs. The third determines the shape of the spectrum and is a string chosen from the following list \{ {\tt monochromatic}, {\tt hm12}, {\tt thermal}, {\tt powerlaw}, {\tt user} \} where {\tt hm12} is the spectral model from \citet{Haardt_12} (HM12) and {\tt user} is a user defined spectrum.  Keyword arguments supply extra information if needed.  For example, the effective temperature in the case of thermal spectra, the powerlaw index in the case of powerlaw spectra, or the redshift in the case of {\tt hm12}. All source classes contain geometry specific normalization routines.  For example, a point source can be normalized to produce a given luminosity in erg~s$^{-1}$ while a plane source can be normalized to produce a given flux in erg~cm$^{-2}$~s$^{-1}$. 

By default, if the range of energies requested spans the hydrogen and helium ionizing thresholds (i.e. if {\tt q\_min} $\le$ 1 and {\tt q\_max} $\ge \nuHeIIth / \nuHIth \approx$ 4) then frequency averaged, or ``grey'', photo-ionization cross sections will be calculated as follows,    
\begin{eqnarray}
\sigmaHI^{\rm grey} &=& 
\frac{ \int_{\nuHIth}^{\nuHeIth} \unu \, \sigmaHI \, d\nu/\nu } 
{ \int_{\nuHIth}^{\nuHeIth} \unu \, d\nu/\nu }
\nonumber 
\\
\sigmaHeI^{\rm grey} &=& 
\frac{ \int_{\nuHeIth}^{\nuHeIIth} \unu \, \sigmaHeI \, d\nu/\nu } 
{ \int_{\nuHeIth}^{\nuHeIIth} \unu \, d\nu/\nu }
\nonumber 
\\
\sigmaHeII^{\rm grey} &=& 
\frac{\int_{\nuHeIIth}^{\nu_{\rm max}} \unu \, \sigmaHeII \, d\nu/\nu} 
{ \int_{\nuHeIIth}^{\nu_{\rm max}} \unu \, d\nu/\nu }
\end{eqnarray}
where $\nu_{\rm max} = {\rm q\_max} \times \nuHIth$, and {\tt q\_max} is one of the arguments provided to the source class. This allows for implicit definitions of grey frequencies and energies, 
\begin{eqnarray}
\label{eq.grey}
\sigma_{\rm X} (\nu_{\rm X}^{\rm grey} ) = \sigma_{\rm X}^{\rm grey}, \quad
E_{\rm X}^{\rm grey} = h \nu_{\rm X}^{\rm grey}
\end{eqnarray}
where ${\rm X}$ is one of \{ HI, HeI, HeII \}.  However, these quantities are calculated only as a convenience to the user.  {\it Unless a monochromatic spectrum is specified, the geometric solvers provided by \rabacus, described in \S \ref{sec.gs}, perform multi-frequency radiative transfer.}  

By default, all spectra are sampled at $\Nnu=128$ photon energies
  logarithmically spaced between {\rm q\_min} and {\rm q\_max}.  The
  variable $\Nnu$ can be passed into the source initialization
  routines to increase or decrease spectral resolution.  All
  photo-ionization and photo-heating rates are calculated by
  performing discretized versions of the integrals in
  Eqs. \ref{eq.pi_rates} and \ref{eq.ph_rates}.  Photo-ionization
  cross-sections are calculated by evaluating the fitting functions of
  \citet{Verner_96} at the $\Nnu$ photon energies.  User
  defined spectra are defined by providing the energy samples and
  spectral shape in tabulated form.  In general, increasing the value
  of $\Nnu$ increases the accuracy of the integrations at the cost of
  longer computations.

\section{Single Zone Solvers}
\label{sec.szs}

The single zone solvers provide equilibrium solutions to Eqs. \ref{eqn.dxHdt},
\ref{eqn.dxHedt}, and \ref{eqn.dTdt}.  Three classes are provided by \rabacus\ to handle three types of equilibrium.

\subsection{Collisional Equilibrium}

The first case we consider is that of collisional equilibrium handled by the class {\tt Solve\_CE}.  In this case, temperatures are fixed, photo-ionization rates are zero, and ionization fractions are found such that collisional ionizations balance recombinations.  In this case, there are closed form solutions for the ionization fractions, 

\begin{eqnarray}
\xHIce = \frac{\RHII}{\RHII + \CHI }, \quad
\xHIIce = \frac{\CHI}{\RHII + \CHI }
\label{eqn.xHce}
\end{eqnarray}

\begin{eqnarray}
\xHeIce &=& \frac{\RHeII \RHeIII}
{\RHeII \RHeIII + \RHeIII \CHeI + \CHeI \CHeII }
\nonumber \\ \nonumber 
\\
\xHeIIce &=& \frac{\RHeIII \CHeI}
{\RHeII \RHeIII + \RHeIII \CHeI + \CHeI \CHeII }
\nonumber 
\nonumber \\ \nonumber 
\\
\xHeIIIce &=& \frac{\CHeI \CHeII}
{\RHeII \RHeIII + \RHeIII \CHeI + \CHeI \CHeII }
\label{eqn.xHece}
\end{eqnarray}
These solutions also determine the minimum possible electron density for a given temperature, $\nelmin = \xHIIce \nH + (\xHeIIce + 2 \xHeIIIce) \nHe$.

\subsection{Photo Collisional Equilibrium}

The second case we consider is that of photo-collisional equilibrium handled by the class {\tt Solve\_PCE}.  In this case, temperatures are fixed, photo-ionization rates are finite, and ionization fractions are found such that collisional and photo ionizations balance recombinations.  In this case, implicit solutions for the ionization fractions are, 

\begin{eqnarray}
\label{eqn.xH_implicit}
\xHI = \frac{\mRHII}{\mRHII + \mIHI}, \quad 
\xHII = \frac{\mIHI}{\mRHII + \mIHI}
\end{eqnarray}

\begin{eqnarray}
\label{eqn.xHe_implicit}
\xHeI &=& \frac{\mRHeII \mRHeIII}
{\mRHeII \mRHeIII + \mRHeIII \mIHeI + \mIHeI \mIHeII}
\nonumber \\ \nonumber 
\\
\xHeII &=& \frac{\mRHeIII \mIHeI}
{\mRHeII \mRHeIII + \mRHeIII \mIHeI + \mIHeI \mIHeII}
\nonumber 
\nonumber \\ \nonumber 
\\
\xHeIII &=& \frac{\mIHeI \mIHeII}
{\mRHeII \mRHeIII + \mRHeIII \mIHeI + \mIHeI \mIHeII}
\end{eqnarray}

They are implicit solutions due to the factors of $\nel$ on the right hand side.  It is possible to eliminate the $\nel$ yielding a univariate fourth order polynomial equation for each ionization fraction, however this yields solutions with thousands of terms.  In addition, the floating point errors involved in numerically evaluating these solutions with fixed precision frequently produce imaginary components.  

In \rabacus\, we use a robust iterative approach.  It involves an initial guess for $\nel$ using a closed form solution for the hydrogen ionization fraction (see \S 6.1).  This allows for the evaluation of all ionization fractions using Eqs. \ref{eqn.xH_implicit} and \ref{eqn.xHe_implicit}.  These are then used to re-calculate $\nel$ using Eq. \ref{eqn.ne} which in turn can be used to re-calculate the ionization fractions.  This process is repeated until the amplitude of changes in $\nel$ drops below a given tolerance.  In most cases, the number of floating point operations for the iterative method will be less than the number required to evaluate the thousands of terms in the closed form solution. In addition, the initial guess for $\nel$ is bounded between a maximum value of $\nel^{\rm max} = \nH + 2 \nHe$ and a minimum value given by collisional ionization equilibrium.

\subsection{Photo Collisional Thermal Equilibrium}

The third case we consider is that of photo-collisional-thermal equilibrium handled by the class {\tt Solve\_PCTE}.  In this case, ionization fractions are found such that collisional and photo ionizations balance recombinations, and a temperature is found such that cooling balances heating.  As described above, we use an iterative technique to find solutions.  First a temperature is guessed then photo-collisional equilibrium is found at that temperature.  Next, heating and cooling rates are calculated using the ionization fractions.  If the heating rate is greater than the cooling rate, the temperature is increased and vice versa.  This procedure is iterated until the heating rate is within a given tolerance of the cooling rate.  The range of temperatures to search can be defined by the user but defaults to $T_{\rm min} = 7.5 \times 10^3$~K and $T_{\rm max} = 10^5$~K.  The initial temperature guess is $\log T = (\log T_{\rm min} + \log T_{\rm max}) / 2$.

\section{Geometric Solver Arguments}
\label{sec.gs}

In the next two sections we present the geometric solver classes available in \rabacus\ and compare our solutions to those in the literature.  These classes take a common set of arguments when instantiated.  The first, {\tt Edges}, defines the gas geometry.  We focus on two geometries: spheres and semi-infinite slabs.  Spherical gas distributions are centered on the origin and discretized into spherical layers separated by infinitesimally thin spherical shells.  In this case, {\tt Edges} is an array of radial distances indicating the position of each shell and typically has zero as the first element and the radius of the sphere as the last element.   Slab gas distributions are discretized into rectilinear layers separated by infinitesimally thin planes.  In this case, {\tt Edges} is an array of distances indicating the position of each plane relative to the surface of the slab and typically has zero as the first element and the thickness of the slab as the last element.  For {\tt N} discrete layers, {\tt Edges} will have {\tt N+1} entries.  The next three arguments, {\tt T}, {\tt nH}, {\tt nHe}, are arrays of length {\tt N} describing the temperature, number density of hydrogen and number density of helium in each layer.

\section{Slab Solvers}
\label{sec.slabs}

\rabacus\ has three slab classes: {\tt SlabPln} which models plane parallel radiation incident from one side; {\tt Slab2Pln} which models plane parallel radiation incident from both sides; and {\tt Slab2Bgnd} which models a slab in a uniform isotropic background.  Our first goal is to validate the \rabacus\ slab solver using a closed form solution in the case of monochromatic radiation and uniform density and temperature. Afterward, we demonstrate the effects of polychromatic spectra, temperature evolution, and recombination radiation.

To begin, we create two plane parallel radiation sources using the {\tt PlaneSource} class.  The first has a spectrum taken from the cosmological UV background model of \citet{Haardt_12} (HM12) at $z=3$ between 1 and 400 Rydbergs.  We then use the neutral hydrogen grey energy (see Eq. \ref{eq.grey}) to create a monochromatic instance of {\tt PlaneSource} and normalize it such that it has the same hydrogen photo-ionization rate as the polychromatic instance of {\tt PlaneSource}.  This results in a source with photons of energy 16.8 eV, a photon flux of $2.31 \times 10^5$~cm$^{-2}$~s$^{-1}$, and a hydrogen photo-ionization rate of $\GHI = 8.27 \times 10^{-13}$~s$^{-1}$.  We then follow the same procedure using the {\tt BackgroundSource} class.  This produces four source objects (a pair of monochromatic sources and a pair of polychromatic sources used to find the grey energies) we can pass to the geometric solvers.

\subsection{Plane Parallel Closed Form Solution}
\label{sec.slab_closed}

In the case of monochromatic plane parallel radiation incident from one side onto a pure hydrogen slab with uniform density and temperature, there is a closed form solution describing the ionization profile if the on-the-spot approximation is used \citep[see the Appendix of][]{Altay_13_urch}, 

\begin{eqnarray}
z \nH \sigma &=& \left( \frac{1}{\xHI^0} - \frac{1}{\xHI} \right) + 
\ln \left[ \frac{\xHI( 1-\xHI^0 )}{\xHI^0 ( 1-\xHI)} \right] 
\nonumber \\
&+& \frac{1}{\xHIce} \ln 
\left[ \frac{\xHI(\xHIce-\xHI^0)}{\xHI^0(\xHIce-\xHI)} \right]
\end{eqnarray} 
where $z$ is depth into the slab, $\sigma$ is the hydrogen photo-ionization cross-section at the monochromatic frequency considered, $\xHI^0$ is the neutral fraction at the surface of the slab, and $\xHI^{\rm ce}$ is the neutral fraction in collisional equilibrium at the fixed temperature of the slab. Note that this is an inverse solution in the sense that it provides depth into the slab as a function of neutral fraction.  However, it allows the full ionization profile to be mapped out by plugging in values of $\xHI$ between the two extremes $\xHI^0$ and $\xHIce$.  Eq. \ref{eqn.xHce} is a closed form solution for $\xHIce$.  There is also a closed form solution for $\xHI^0$, 
\begin{eqnarray}
\xHI^0 = \frac{-B - (B^2 - 4AC)^{1/2}}{2A}
\end{eqnarray}  
The coefficients in this quadratic solution are,
\begin{eqnarray}
A &=& (\RHII + \CHI) \nH
\nonumber \\
B &=& -[ \GHI^0 + \RHII \nH + A (1+y) ]
\nonumber \\
C &=& \RHII \nH (1+y)
\end{eqnarray}  
where $y$ represents the contribution to free electrons from elements other than hydrogen, $ \nel = (\xHII + y) \nH $, and $\GHI^0$ is the photo-ionization rate at the surface of the slab \citep{Altay_13_urch}.

To generate an appropriate slab model in \rabacus, we use the {\tt Cosmology} class  to calculate the critical density of hydrogen at $z=3$ using the Planck cosmological parameters \citep{Planck_Cosmo_13} and then calculate the Jeans length (assuming a mean molecular weight $\mu=1$) for gas at $10^4$~K and $\Delta=2200$ times the critical density. The value for $\Delta$ was arbitrarily chosen to produce a small neutral region in the slab and results in a length of $L = 4.17 $~kpc and number densities of $\nH = 8.61 \times 10^{-3}$~cm$^{-3}$ and $\nHe = 7.09 \times 10^{-4}$~cm$^{-3}$.

\begin{figure}
\begin{center}
\includegraphics[width=0.47\textwidth]
{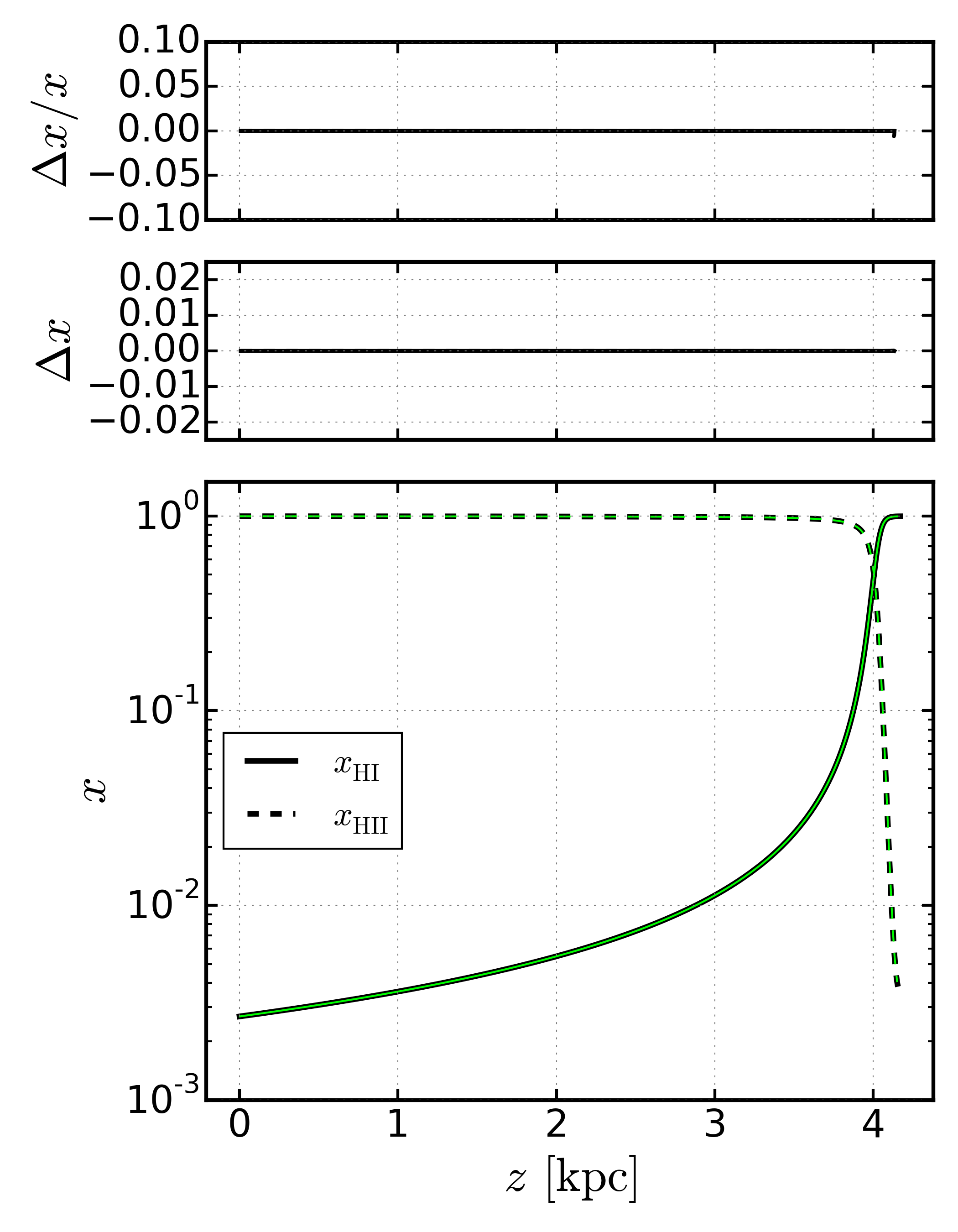}
\caption{Ionization fractions for the closed form solution test
  described in section \ref{sec.slab_closed}.  In the lower panel we
  show ionization fractions from \rabacus\ (thick lines) and the
  closed form solution (thin lines).  In the middle panel we show the
  absolute difference in ionization fractions $x^{\rm ana}$ -
  \rabacus.  In the top panel we show the relative difference in
  ionization fractions ( $x^{\rm ana}$ - \rabacus\ ) / \rabacus.  In
  the bottom panel, the \rabacus\ solution is shown in black while the
  analytic solution is shown in green. }
\label{fig.slab_closed}
\end{center}
\end{figure}

In Fig. \ref{fig.slab_closed} we show the closed form ($\xHI^{\rm ana}$, $\xHII^{\rm ana}$) and \rabacus\ solutions.  We generated the \rabacus\ solution by passing the monochromatic plane source and the slab description to the {\tt SlabPln} class.  The maximum relative and absolute difference between the two solutions occurs near the ionization front but is always less than $7 \times 10^{-3}$ and two orders of magnitude smaller than that over most of the length of the slab. This indicates that the basic machinery of the \rabacus\ solver is correct. 


\subsection{Isotropic Background, Recombination Radiation}

In the previous section both the gas geometry and the radiation transport were one dimensional.  In this section we discuss isotropic backgrounds and diffuse recombination radiation both of which introduce angular variables into the problem.  
The {\tt Slab2Bgnd} class models an isotropic background incident onto a planar gas distribution.  Here we describe the equations that are solved by \rabacus\ to account for both the background radiation and radiation from recombinations.  We also note that recombination radiation can optionally be treated the same way in the {\tt SlabPln} and {\tt Slab2Pln} classes. 

\begin{figure}
\begin{center}
\includegraphics[width=0.47\textwidth]
{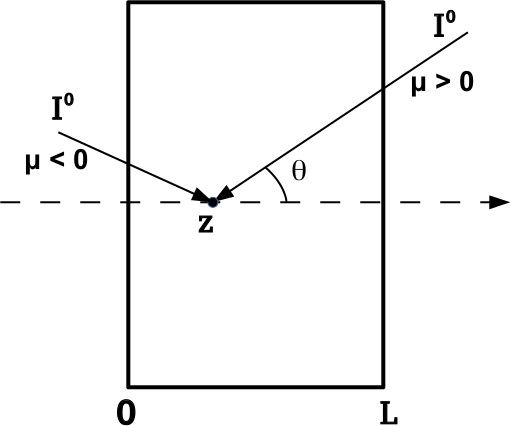}
\caption{ The geometry used in the {\tt SlabBgnd} class. }
\label{fig.slab_geometry}
\end{center}
\end{figure}

The angle averaged specific intensity at any depth $z$ (see Fig. \ref{fig.slab_geometry}) into the slab is, 
\begin{equation}
\Jnu(z) = \frac{1}{4 \pi} \int_0^{2\pi} d\phi \int_{-1}^{1} d\mu \, \Inu(z,\mu)
\end{equation}
When considering a point at depth $z$, it is convenient to decompose the rays into those with negative z-components ($\Inu^-$) for which $\mu \equiv \cos \theta > 0$ and those with positive z-components ($\Inu^+$) for which $\mu < 0$.  In addition, we will decompose the radiation into a component from the background itself (${\rm src}$) and a component from radiative recombinations in the slab (${\rm rec}$).  Using this decomposition we can write, 
\begin{eqnarray}
  \Inu^{\rm src+}(z,\mu) &=& \Inu^0 \exp \left[ -\tau(z) / |\mu| \right]
  \\
  \Inu^{\rm src-}(z,\mu) &=& \Inu^0 \exp \left[ -\dtau{L}{z} / \mu \right]
  \\
  \Inu^{\rm rec+}(z,\mu) &=& \int_0^z \frac{dz'}{|\mu|} \jnu(z') 
                           \exp \left[ -\dtau{z}{z'} / |\mu| \right] 
  \\
  \Inu^{\rm rec-}(z,\mu) &=& \int_z^L \frac{dz'}{\mu} \jnu(z') 
                           \exp \left[ -\dtau{z'}{z} / \mu \right]
\end{eqnarray}  
where $\dtau{a}{b} \equiv \tau(a) - \tau(b)$, $\tau(z')$ is the optical depth between $z=0$ and $z=z'$ along a path perpendicular to the surface of the slab, $\jnu$ is the emission coefficient for recombination radiation, and $\Inu^0$ is the un-attenuated isotropic background.  
The angle averaged specific intensity can now be written as, 
\begin{equation}
  \Jnu(z) = \frac{1}{2} 
  \left[ \int_{-1}^{0} \left( \Inu^{\rm src+} + \Inu^{\rm rec+} \right) d\mu  +
    \int_{0}^{1} \left( \Inu^{\rm src-} + \Inu^{\rm rec-} \right) d\mu \right] 
\end{equation}
The angular integrals can be expressed in terms of the exponential integrals
$E_1$ and $E_2$, 
\begin{eqnarray}
\Jnu^{\rm src+}(z) &=& \frac{1}{2} \int_{-1}^{0} \Inu^{\rm src+} d\mu = \frac{\Inu^0}{2} E_2 \left[ \tau(z) \right]
\\ 
\Jnu^{\rm src-}(z) &=& \frac{1}{2} \int_{0}^{1} \Inu^{\rm src-} d\mu =  
\frac{\Inu^0}{2} E_2 \left[ \dtau{L}{z} \right] 
\end{eqnarray}


\begin{equation}
\Jnu^{\rm rec+}(z) = 
\frac{1}{2} \int_{-1}^{0} \Inu^{\rm rec+} d\mu = 
\frac{1}{2} \int_0^z dz' \jnu(z') E_1 \left( \dtau{z}{z'} \right)
\end{equation}

\begin{equation}
\Jnu^{\rm rec-}(z) = 
\frac{1}{2} \int_{0}^{1} \Inu^{\rm rec-} d\mu = 
\frac{1}{2} \int_z^L dz' \jnu(z') E_1 \left( \dtau{z'}{z} \right)
\end{equation}
Combining all terms we get, 
\begin{eqnarray}
\Jnu(z) &=& \frac{\Inu^0}{2} \left\{ 
E_2 \left[ \tau(z) \right] + E_2 \left[ \dtau{L}{z} \right] \right\} +
\nonumber \\
&& \frac{1}{2} \int_0^L dz' \jnu(z') E_1 \left( |\dtau{z'}{z}| \right)
\end{eqnarray}
We account for recombination radiation from hydrogen and helium with monochromatic emission at the appropriate ionization threshold and calculate the emission coefficients as, 
\begin{eqnarray}
\label{eq.jcoefH}
j_{\rm HII} &=& \frac{\RHII^1 \nH \xHII \nel}{4 \pi}
\\
\label{eq.jcoefHe}
j_{\rm HeII} = \frac{\RHeII^1 \nHe \xHeII \nel}{4 \pi}, && 
j_{\rm HeIII} = \frac{\RHeIII^1 \nHe \xHeIII \nel}{4 \pi} 
\end{eqnarray}
where the $R^1_{_X}$ represent recombinations to the ground state of
each ion.  In practice, this rate is calculated by subtracting the
case B fit from the case A fit. This neglects some helium
recombinations that produce ionizing photons \citep[see for example \S 3.1 in][]{Altay_08} but makes for a more straightforward
comparison to models which use case B rates only.  Our treatment is similar to the models used in \citet{CAFG_09} and \citet{Haardt_12} except that we include the effects of collisional ionization.


\subsection{Isolating Physical Processes}
\label{sec.slab2_full}

\begin{figure*}
\begin{center}
\includegraphics[width=0.94\textwidth]
{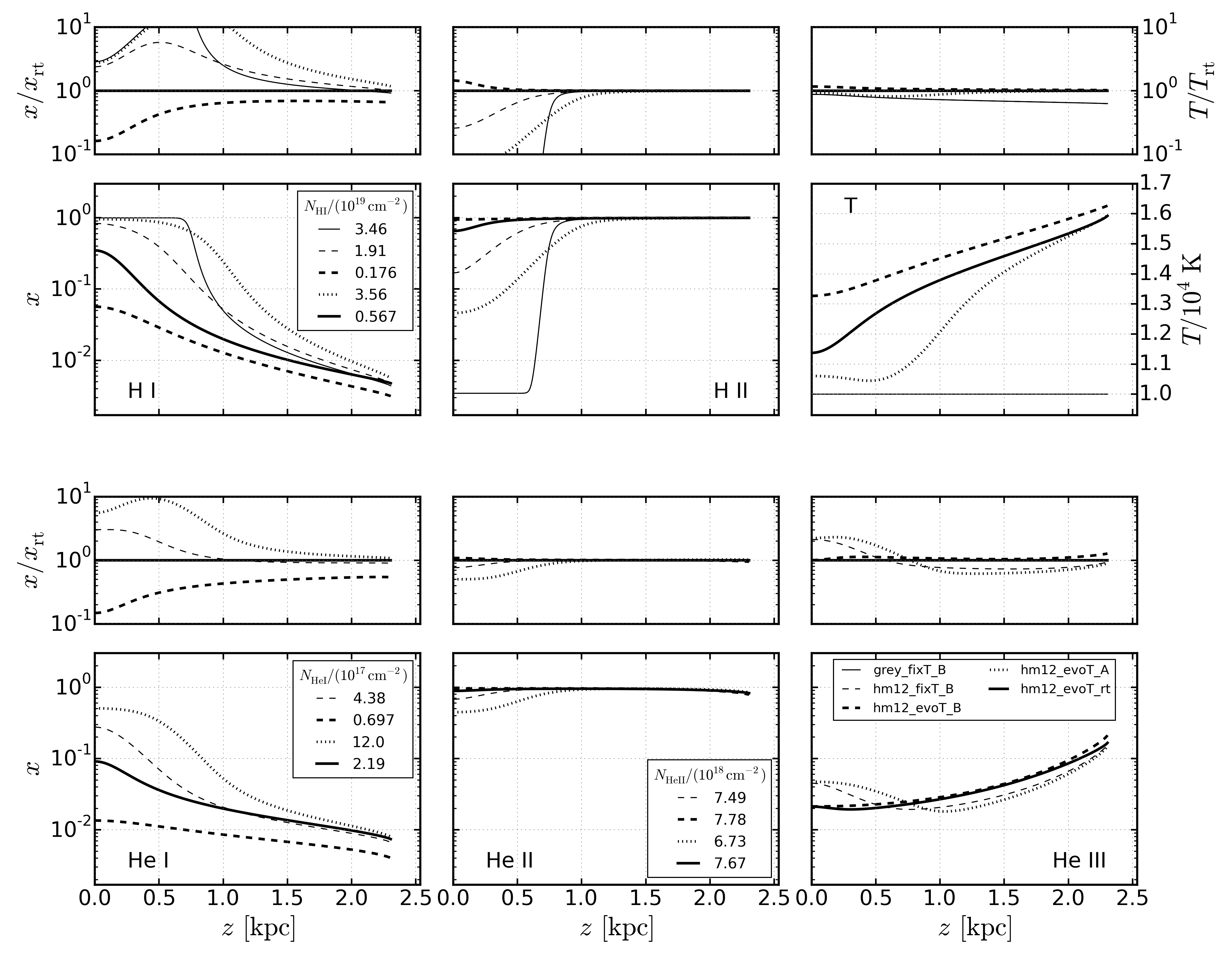}
\caption{ Ionization fraction and temperature profiles in five simple models of a Lyman limit system.  Each panel shows the ionization fraction of one ion or the temperature.  The five models considered are\protect{:} 1) {\bf grey\_fixT\_B} (thin solid) in which the grey approximation to the HM12 spectrum is used, the temperature is kept fixed, and recombinations are treated using case B rates, 2) {\bf hm12\_fixT\_B} (thin dashed) in which the full HM12 spectrum is used, the temperature is kept fixed, and recombinations are treated using using case B rates, 3) {\bf hm12\_evoT\_B} (thick dashed) in which the full HM12 spectrum is used, equilibrium temperatures are calculated, and recombinations are treated using case B rates, 4) {\bf hm12\_evoT\_A} (dotted) as in {\bf hm12\_evoT\_B} but with case A recombination rates (i.e. recombination radiation is ignored) and 5) {\bf hm12\_evoT\_rt} (thick solid) in which the full HM12 spectrum is used, equilibrium temperatures are calculated, recombination rates are case A, and recombinations are treated by calculating the transfer of  radiation through the system.  Above each profile we show the ratio of each model's prediction to those of {\bf hm12\_evoT\_rt} (i.e. dex of difference in the profile panels).  The scale in the top panels is a factor of ten in either direction. }
\label{fig.slab2_full_x}
\end{center}
\end{figure*}

Having verified the basic functionality of \rabacus\  against a closed form solution, we now show the results of including more physical processes.  All geometric solvers in \rabacus\ include the ability to treat  polychromatic spectra, helium ionization, thermal balance, and the transfer of recombination radiation.  We now use the slab geometry to produce five models which add these physical processes one at a time.  For all of these examples we will use the {\tt Slab2Bgnd} class which models a slab in an isotropic background of radiation.  In general, plane parallel radiation will penetrate further into a slab than isotropic radiation.  Therefore, we use a lower overdensity of $\Delta = 1800$ in this test.  This leads to a slab 
of length $L = 4.61 $~kpc and number densities $\nH = 7.04 \times
10^{-3}$~cm$^{-3}$ and $\nHe = 5.80 \times 10^{-4}$~cm$^{-3}$.
Although the temperature, and hence the Jeans length, will deviate
from $10^4$ K when we solve for thermal balance, we keep the slab size
fixed for these tests.  This results in a neutral hydrogen column
density of $\NHI = 5.67 \times 10^{18}$~cm$^{-2}$ for our most
realistic model.  These volume and column densities are typical for
Lyman limit absorption systems.  The five models we examine are,

\begin{description}
\item[grey\_fixT\_B] monochromatic spectrum, fixed temperature, treats recombination photons by using case B rates
\item[hm12\_fixT\_B] adds full HM12 spectrum as opposed to the monochromatic grey approximation
\item[hm12\_evoT\_B] adds thermal balance as opposed to fixed temperatures
\item[hm12\_evoT\_A] same as {\bf hm12\_evoT\_B} but with case A rates
\item[hm12\_evoT\_rt] uses case A recombination rates but adds the transfer of recombination photons from hydrogen and helium emitted isotropically from each layer
\end{description}


In Fig. \ref{fig.slab2_full_x} we show ionization and temperature profiles for all five models.  Note that we have taken advantage of the symmetric geometry and only show half the slab.  Above each profile panel we show the ratio of each model's predictions to those of {\bf hm12\_evoT\_rt} (i.e. dex of difference in the bottom panel).  

By comparing models {\bf grey\_fixT\_B} and {\bf hm12\_fixT\_B} we can isolate the effect of using the grey approximation. The monochromatic model is characterized by a fully neutral hydrogen core of approximately 1.5 kpc and sharp edges.  Using the polychromatic spectrum smooths the hydrogen ionization profile and reduces the neutral hydrogen column density through the slab, $\NHI$, by a factor of 1.8.  The monochromatic spectrum does not contain any helium ionizing photons and so a comparison between the helium ionization profiles is not well motivated.  

By comparing models {\bf hm12\_fixT\_B} and {\bf hm12\_evoT\_B} we can isolate the effect of self-consistently calculating the temperature structure. The equilibrium temperature is greater than 10$^4$~K at all depths in the slab and peaks at 1.6 $\times 10^4$~K at the edge.  This produces both higher collisional ionization rates and lower recombination rates leading to fewer neutral hydrogen and helium atoms.  This effect is compounded by the additional photons that can penetrate into the more ionized slab. Calculating the equilibrium temperatures lowers $\NHI$ by a factor of 11 and $\NHeI$ by a factor of 6.3 while $\NHeII$ is only affected at the five percent level. 

By comparing models {\bf hm12\_evoT\_B} and {\bf hm12\_evoT\_A} we get a sense of the maximum variation possible due to recombination photons.  In the case B model, recombination photons are emitted and absorbed in the same layer, while in the case A model all recombination photons escape the system.  In model {\bf hm12\_evoT\_rt} we calculate the transfer of recombination photons isotropically emitted from each layer.  The rt and case A solutions agree at the optically thin surface of the slab.  As depth increases, the rt solution diverges from the case A solution but always remains between the case A and case B solutions.  This translates into a factor of 3.2 increase in both $\NHI$ and $\NHeI$ relative to case B.  These changes are larger than those that result from using a grey spectrum as opposed to a polychromatic one.  The columns of singly and doubly ionized helium are only affected at the few percent level. 

The fact that the rt solution does not converge to the case B result at any depth distinguishes the ionized slab model from models like the Str\"{o}mgren sphere (see \S \ref{sec.strom_spheres}).  The major difference is that Str\"{o}mgren spheres necessarily have a neutral region enclosing the source of photons.  Conservation of photons requires that all solutions agree at radii large enough where all photons have been absorbed.  However, when the slab is not fully neutral at its center (as will be the case for Lyman limit systems as well as lower column density systems), the rt solution will differ from the case B solution at all points in the slab. In summary, the behavior of the most realistic solution here is not captured by either case A or case B approximations.  In addition, the total column density in neutral hydrogen and helium differs between the models by at least a factor of 3 and at most a factor of 11.


\section{Str\"{o}mgren Sphere Solver}
\label{sec.strom_spheres}

\rabacus\ has two sphere classes: {\tt SphereStromgren} which models a point source at the center of a sphere and {\tt SphereBgnd} which models a sphere placed in a uniform and isotropic background.  There is no closed form solution that describes the ionization profile in either case, however solutions are readily found using numerical integration.  In this section and the next we present the solutions of \rabacus\ and compare them to the results of the photo-ionization code \cloudy\ version 13.03 last described in \cite{Ferland_13}.  

We do not expect perfect agreement between the two codes for several reasons.  First, \cloudy\ makes use of a more detailed physical model including multi-level atoms and a large number of line transfer processes which are not included in \rabacus.  Second, the convergence criteria for \cloudy\ allow for a deviation from unity in the sum of the ionization fractions of a given element (for example $\xHI + \xHII$) at the level of $10^{-4}$.  Third, the  atomic rates that determine the exact ionization and temperature structure in a parcel of gas are only known to a few significant digits at best and are not identical in the two codes.  This is why we have tested \rabacus\ in previous sections against closed form analytic solutions.  It follows that differences between \cloudy\ and \rabacus\ should be considered a measure of uncertainty as opposed to an error in either code, both of which have been verified against analytic solutions.

The transfer of radiation from a central point source (ignoring recombination radiation for the moment) is a one-dimensional problem.  This means the same techniques that we used in the case of plane parallel radiation incident on a slab geometry can be applied to the sphere as long as we include the inverse square geometric dilution of the radiation.  In the following sections we verify the spherical solver in \rabacus\ using a simple test problem and then study the effects of temperature evolution, helium, and recombination radiation.

\subsection{ Pure Hydrogen, Fixed Temperature }

\begin{figure}
\begin{center}
\includegraphics[width=0.47\textwidth]
{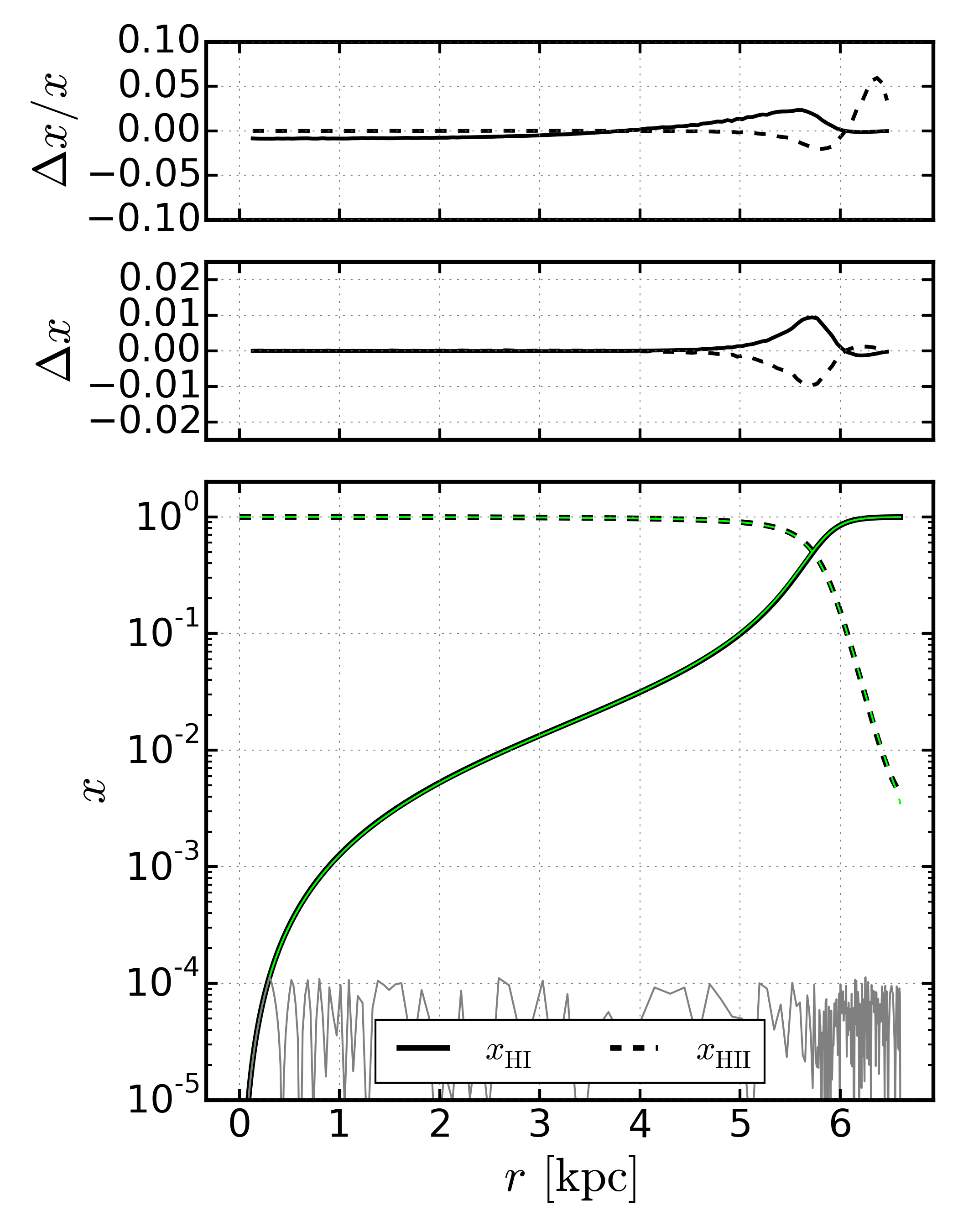}
\caption{Ionization fractions for Test 1 of \protect \cite{Iliev_06}
  (with 15 eV photons).  In the lower panel we show ionization
  fractions from \rabacus\ (thick lines) and \cloudy\ (thin lines).
  The deviation of the sum of the ionization fractions from unity
  ($\delta_{\rm H} = |1-\xHI-\xHII|$) in the \cloudy\ model is shown
  as a thin grey line.   In the middle panel we show the absolute
  difference in ionization fractions \cloudy\ - \rabacus.  In the top
  panel we show the relative difference in ionization fractions (
  \cloudy\ - \rabacus\ ) / \rabacus.  In the bottom panel, the
  \rabacus\ solution is shown in black and the \cloudy\ solution is
  shown in green. }
\label{fig.iliev_test1_x}
\end{center}
\end{figure}

We begin by examining Test 1 described in \citet{Iliev_06} which considers a pure hydrogen sphere of radius $6.6$~kpc with uniform density, $\nH = 10^{-3}$~cm$^{-3}$, and temperature, $T=10^4$~K.  The source has a photon luminosity of $L_{\rm n} = 5.0 \times 10^{48}$ per second.  The test specifies a monochromatic spectrum with all photons having energy $13.6$ eV.  However, all spectra in \cloudy\ have a finite width and so creating a monochromatic spectrum at the Lyman limit produces some non-hydrogen ionizing photons.  In order to make a clean comparison to \rabacus\ we instead use a monochromatic spectrum at 1.1 Ry or 15 eV.  Recombination photons are treated by using the case B recombination rate (i.e. the on-the-spot approximation).  In Figure \ref{fig.iliev_test1_x} we present ionization profiles from both codes.  The relative difference between the two solutions is typically less than one percent but can be a few percent near the ionization front.

\subsection{ Temperature Evolution }

\begin{figure}
\begin{center}
\includegraphics[width=0.47\textwidth]
{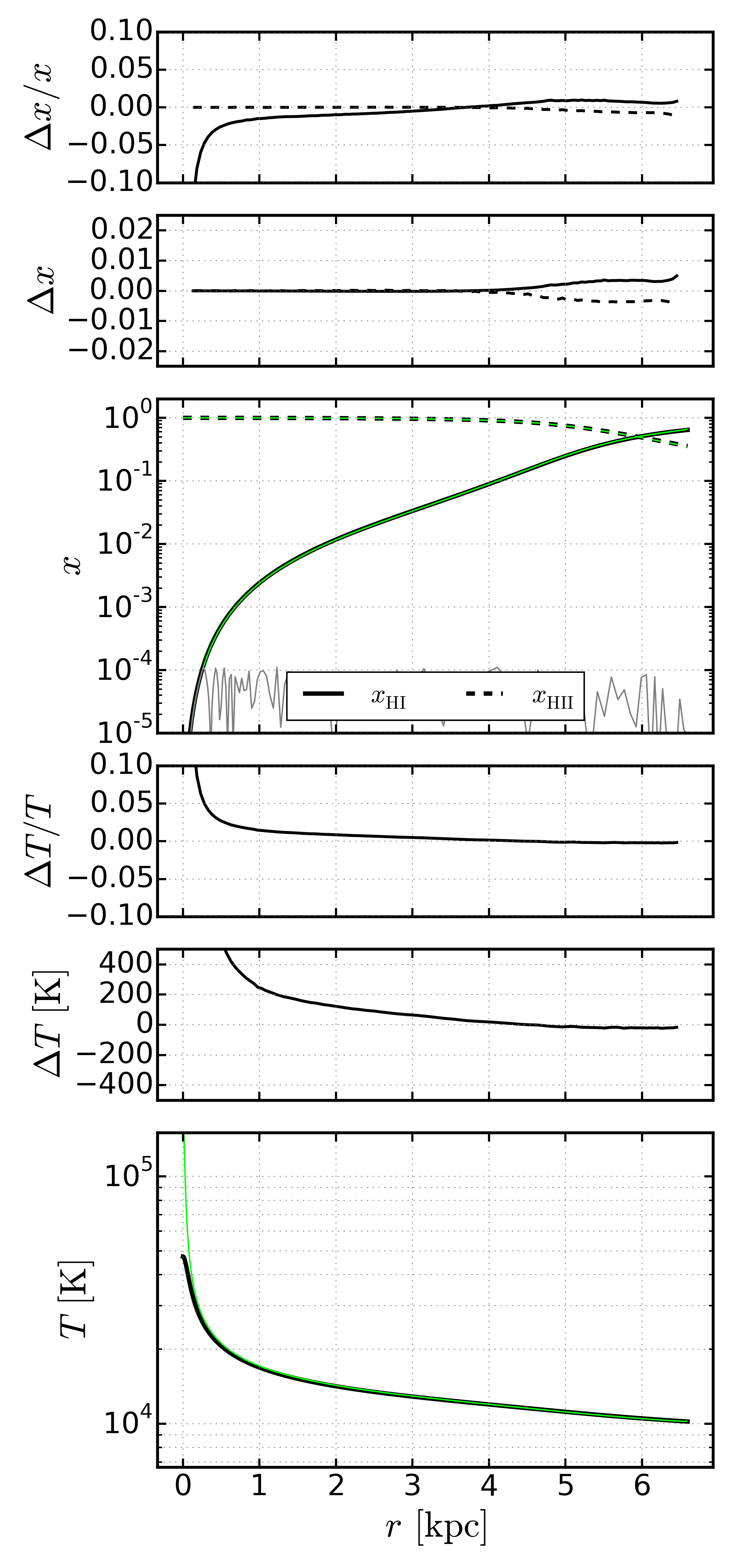}
\caption{Ionization and temperature profiles for Test 2 of \protect
  \cite{Iliev_06}.  In the upper three panels we show ionization
  fractions from \rabacus\ (thick black lines) and \cloudy\ (thin
  green lines).  The deviation of the sum of the ionization fractions
  from unity ($\delta_{\rm H} = |1-\xHI-\xHII|$) in the \cloudy\ model
  is shown as a thin grey line.  The top two panels show the absolute
  difference in ionization fractions, \cloudy\ - \rabacus, and the
  relative difference in ionization fractions, ( \cloudy\ -
  \rabacus\ ) / \rabacus.  The bottom three panels show the same thing 
  for temperature. The difference between
  the two codes in the temperature profile panel is only significant for
  the innermost quarter kpc.  These differences coincide with the
  radius at which $\xHI$ approaches $\delta_{\rm H}$ in the \cloudy\ model.  }
\label{fig.iliev_test2}
\end{center}
\end{figure}

The second test from \citet{Iliev_06} allows for temperature evolution and includes a black body spectrum with effective temperature $T_{\rm eff} = 10^5$~K. We only consider photons with energy between 1 and 10 Rydbergs and normalize the spectrum such that the source emits $L_{\rm n} = 5 \times 10^{48}$ photons per second.  The temperature profile of the sphere is determined by balancing photo-heating and radiative cooling processes.  We present the solutions from \rabacus\ and \cloudy\ in Fig. \ref{fig.iliev_test2}.  The relative difference in ionization fractions and temperature between the two codes for this test is on the order of a few percent except in the innermost quarter kpc.  These differences coincide with the radius at which $\xHI$ approaches the magnitude of the deviation from unity of the ionization fractions, $\delta_{\rm H} = |1-\xHI-\xHII|$, in the \cloudy\ model.

\subsection{ Inclusion of Helium }

\begin{figure}
\begin{center}
\includegraphics[width=0.47\textwidth]
{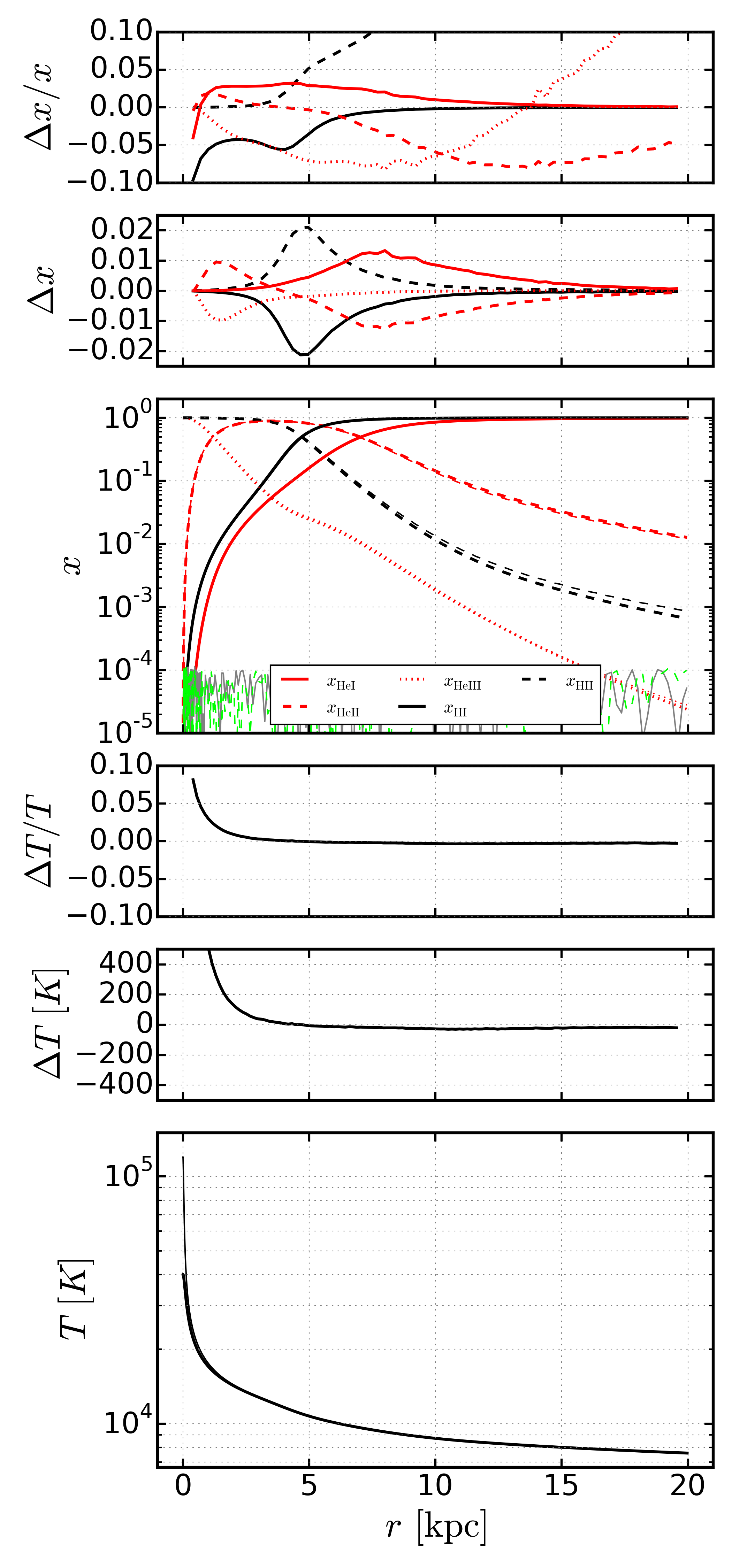}
\caption{Ionization and temperature profiles for Test 1 A of \protect \cite{Friedrich_12} with the addition of temperature evolution.  In the upper three panels we show ionization fractions from \rabacus\ (thick lines) and \cloudy\ (thin lines).  The deviation of the sum of the ionization fractions from unity ($\delta_{\rm H} = |1-\xHI-\xHII|$ and $\delta_{\rm He} = |1-\xHeI-\xHeII-\xHeIII|$) in the \cloudy\ model are shown as thin grey and green lines respectively.  The top two panels show the absolute difference in ionization fractions, \cloudy\ - \rabacus, and the relative difference in ionization fractions, ( \cloudy\ - \rabacus\ ) / \rabacus.  The lines from the two codes are only distinguishable for the ionized species at large radii.  The bottom three panels show the same thing for temperature. The difference between the two codes in the temperature profile panel is only visible at small radii. }
\label{fig.friedrich_test1at}
\end{center}
\end{figure}

\citet{Friedrich_12} extend the spherically symmetric tests described in \citet{Iliev_06} to include a uniform helium density of $\nHe = 8.7 \times 10^{-5}$ cm$^{-3}$ and increase the radius of the sphere to 20 kpc.  To verify the treatment of helium in \rabacus\, we use a $10^5$~K blackbody spectrum, case A recombination rates (i.e. ignore recombination radiation), and allow temperatures to vary as in the previous test.  We consider photons with energy between 1 and 10 Rydbergs and normalize the spectrum such that the source emits $L_{\rm n} = 5 \times 10^{48}$ photons per second.  Except for the addition of temperature evolution, this is the same setup as Test 1 A in \citet{Friedrich_12}.   We use both \rabacus\ and \cloudy\ to generate solutions and present the ionization and temperature profiles in Fig. \ref{fig.friedrich_test1at}.  At small radii we see the same effect as in the previous test due to the neutral fractions approaching $\delta_{\rm H} = |1-\xHI-\xHII|$ and $\delta_{\rm He} = |1-\xHeI-\xHeII-\xHeIII|$.  At larger radii, the relative differences in the hydrogen and helium neutral fractions are always less than 5\%.  In the case of the ionized species, the relative differences between ionization fractions in \rabacus\ and \cloudy\ is less than 10\% for most radii.  However, for ionized hydrogen and doubly ionized helium, the differences can be greater.  It is hard to determine the exact cause of these differences, but the most likely explanations are differences in the atomic rates used in the two codes and line processes in \cloudy\ that are not included in \rabacus. The absolute difference between ionization fractions in the two codes is always less than 0.02.

\subsection{ Hydrogen Recombination Radiation }

In the previous Str\"{o}mgren sphere examples, we have either ignored recombination radiation (i.e. used case A rates) or treated it using the on-the-spot approximation (i.e. used case B rates).   A more accurate treatment would allow photons produced via recombinations to travel some distance in the system before producing a photo-ionization.  The outward only approximation is commonly used in analytic treatments of Str\"{o}mgren spheres \citep[e.g.][]{Ritzerveld_05}.  In this approximation, all recombination photons are assumed to be emitted in the outward radial direction.  This is equivalent to a situation in which half the photons are emitted radially inward and half radially outward {\it if} the optical depth seen by recombination photons interior to the emitting layer is zero.  \citet{Raicevic_14} examine the effects of isotropically transporting recombination radiation in the Str\"{o}mgren sphere problem using the same physical setup as in Test 1 of \citet{Iliev_06}.  \rabacus\ includes options to treat the isotropic emission of recombination radiation as well as options to use the outward only approximation.  In this section, we present solutions produced by \rabacus\ with isotropic transport of recombination radiation.     

\begin{figure}
\begin{center}
\includegraphics[width=0.47\textwidth]
{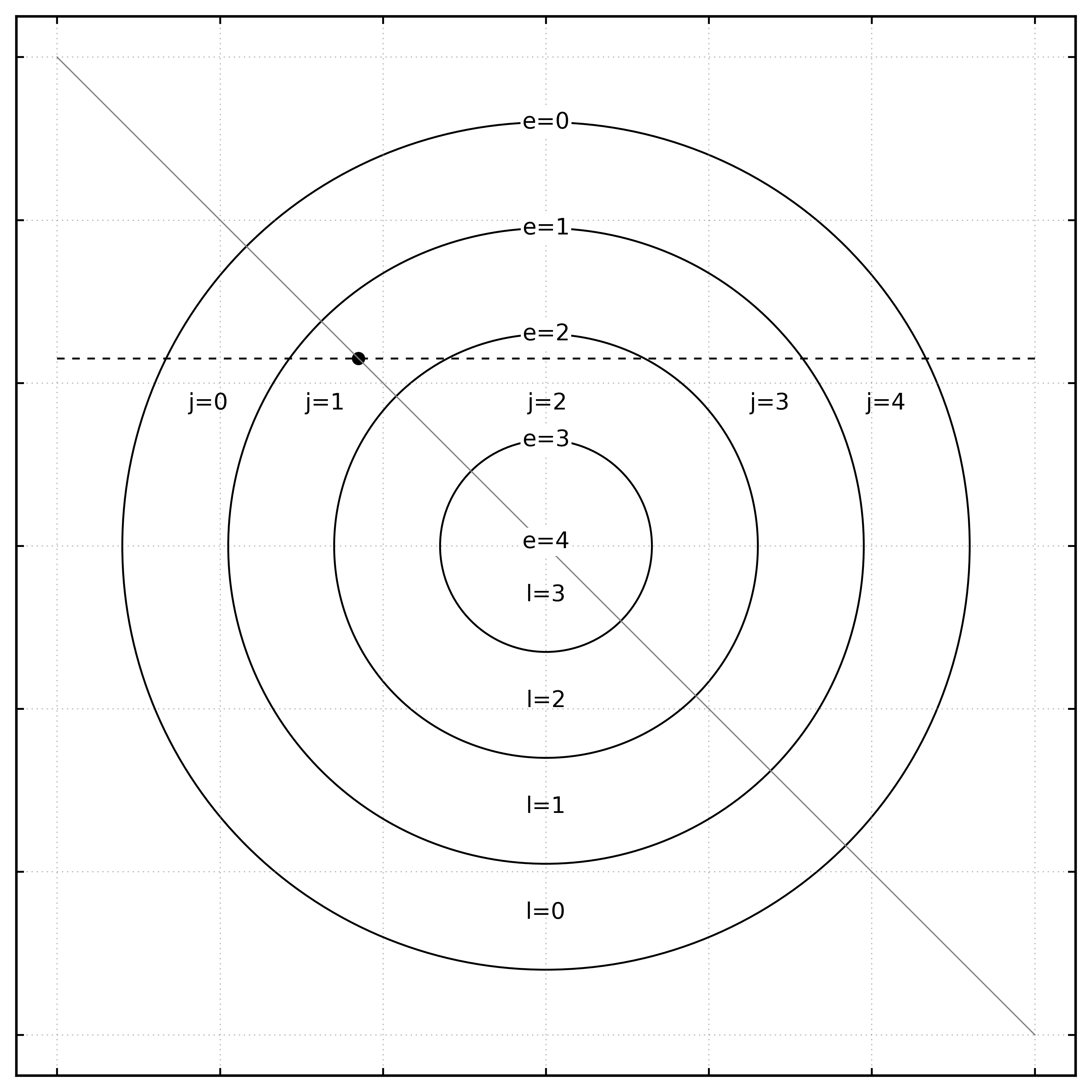}
\caption{Illustration of ray tracing in spherical geometries.  Because
  of symmetry, all rays (dashed line) can be rotated such that they are
  perpendicular to the vertical axis.  We indicate the origin of the
  rays with a dot in the center of the $l=1$ layer and the radial
  direction with a grey solid line.  In the calculation of path length
  $ds$ through each layer we label the edges (from large to small
  radii) with the integer $e$, the layers (from large to small radii)
  with the integer $l$ and the ray segments (from left to right) with
  the integer $j$. }
\label{fig.sphere_geo}
\end{center}
\end{figure}

\begin{figure}
\begin{center}
\includegraphics[width=0.47\textwidth]
{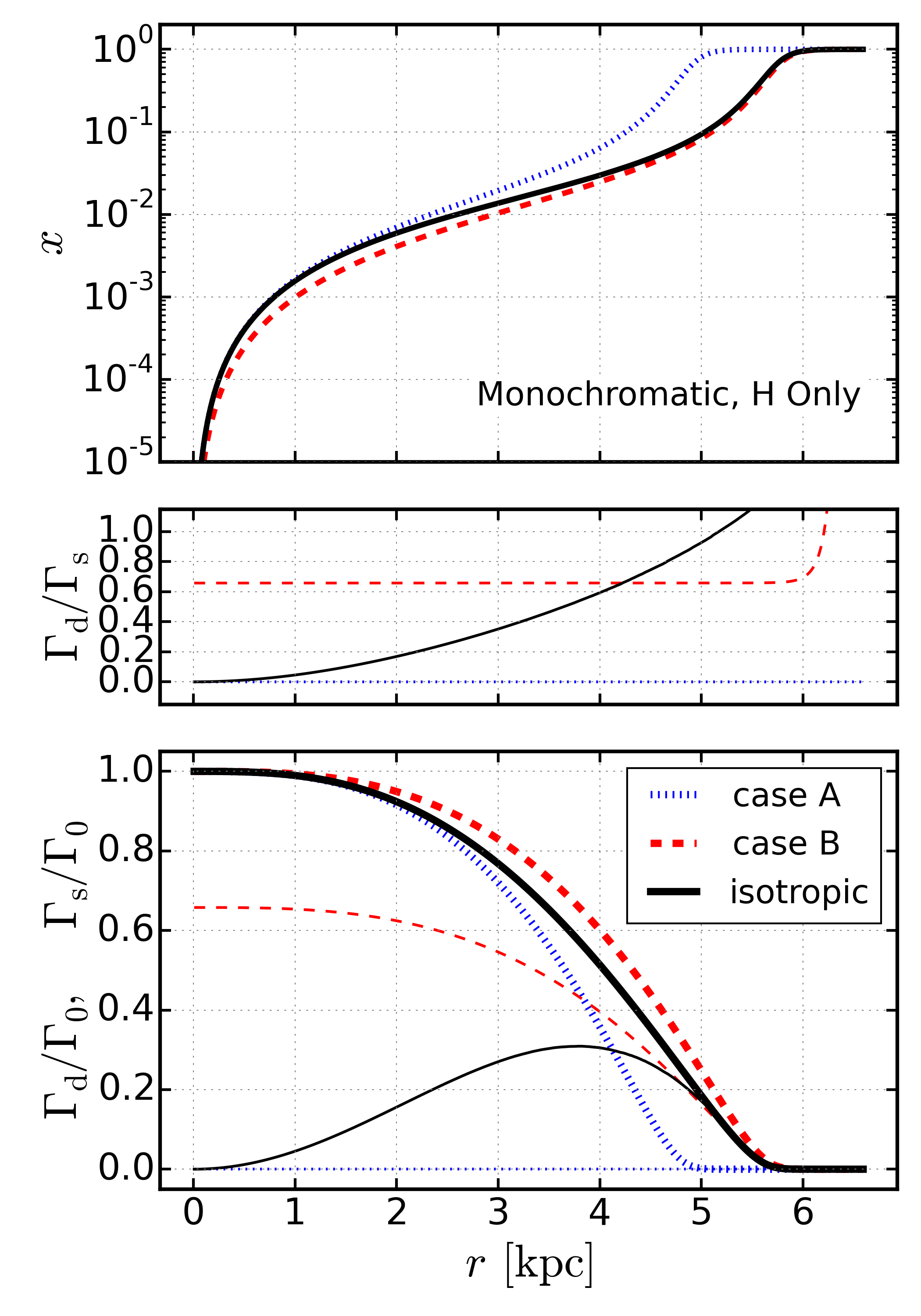}
\caption{Results for the test described in \protect \citet{Raicevic_14}.  In the upper panel we show the neutral hydrogen fraction resulting from using case A rates (ignoring recombination radiation), using case B rates (using the on-the-spot approximation) and from transporting recombination radiation isotropically.  In the lower panel we show photo-ionization rates due to the central source (thick lines) and due to recombination radiation (thin lines) all with geometric dilution scaled out.  In the middle panel we show the ratio of diffuse to central source rates.  }
\label{fig.raicevic_H_mono}
\end{center}
\end{figure}

When radiation is not constrained to move in the radial direction, one
must account for the angular dependence of the optical depth when
calculating photo-ionization and photo-heating rates.  In spherical
geometries this angular dependence is restricted to a dependence on the
polar angle $\mu = \cos \theta$.  To calculate the transfer of
recombination radiation, one must calculate the optical depth through
the sphere along $N_{\mu}$ directions for every layer.  Each ray
begins at the center of a layer and the directions, $\mu_i$ are chosen  
using Gauss-Legendre quadrature.  The number of directions can be specified by the user but defaults to $N_{\mu}=32$.  We found that increasing or decreasing the number of angles by a factor of 2 had no effect on the tests presented here, however users should check convergence for each problem. 

With the proper rotation, all rays can be made perpendicular to the
vertical axis.  In Fig. \ref{fig.sphere_geo} we show two rays that
originate from the layer $l=1$.  The first (extending to the left)
represents a direction with $\mu_i = \cos \theta > 0$ while the second
(extending to the right) represents a direction with $\mu_i = \cos
\theta < 0$.  We label the edges of the sphere (from large to small
radii) with the integer $e$, the layers of the sphere (from large to
small radii) with the integer $l$, and the segments of the ray (from
left to right) with the integer $j$.  Given a layer of origin and a
direction $\mu_i$, we first calculate the impact parameter $b$ and the
layer of closest approach $l_b$ (e.g. $l_b=2$ in Fig. \ref{fig.sphere_geo}).
In addition we label the radius of each edge with the function
$r_e(i)$.  Given this information, the path length $ds(j)$
through segment $j$ can be calculated as follows, 
\begin{equation}
ds(j) = \left\{ \,
\begin{IEEEeqnarraybox}[][c]{l?s}
\IEEEstrut
\sqrt{r^2_e(j+1)-b^2} - \sqrt{r^2_e(j)-b^2} & if $j<l_b$, \\
2 \sqrt{r^2_e(j)-b^2} & if $j=l_b$, \\
\sqrt{r^2_e(k+1)-b^2} - \sqrt{r^2_e(k)-b^2} & if $j>l_b$.
\IEEEstrut
\end{IEEEeqnarraybox}
\right. 
\label{eq:ds}
\end{equation}
\begin{equation}
k = l_b - \abs{l_b-j}
\end{equation}
When modeling an isotropic background external to the sphere, one
adds the optical depth from each ray segment $j$ between the ray
origin and the surface of the sphere.  This is then used to attenuate
the incoming radiation.  When modeling recombination radiation,
emission from each layer is attenuated using the appropriate optical
depth segments.

In the upper panel of Fig. \ref{fig.raicevic_H_mono} we show hydrogen ionization profiles when using case A recombination rates (i.e. ignoring recombination radiation), when using case B recombination rates (i.e. the on-the-spot approximation), and when transporting recombination radiation isotropically.  As expected, the case A result produces the smallest ionized region.  The case B result assumes that each recombination produces a local ionization and so produces a larger ionized region.  When recombination photons are emitted isotropically, they travel through the low optical depths at small radii without being absorbed and the solution tracks the case A model.  However, all photons are eventually absorbed in the Str{\"o}mgren sphere problem and so the final radius of the Str{\"o}mgren sphere must be the same as in the case B approximation due to photon conservation. This result does not hold for slab geometries which are optically thin at their center (see \S \ref{sec.slab2_full}). 

An instance of the {\tt SphereStromgren} class has attributes that store photo-ionization rates due to both the central source and to diffuse recombination radiation.  In the lower panels of Fig. \ref{fig.raicevic_H_mono} we show how these  scale with radius.  The diffuse field for the on-the-spot approximation is calculated using what we term the effective photo-ionization rate.  In addition, to tracking the source and diffuse photo-ionization rates, \rabacus\ will define an effective photo-ionization rate, $\Gamma^{\rm eff}$, for each ionized species if recombination photons are not being transported.  The effective rates are defined for each layer to be those that would result in the same ionization fractions {\it if} case A recombination rates were used.  By definition the effective photo-ionization rate is equal to the actual photo-ionization rate if case A rates are used. However, if case B rates are used, subtracting the actual photo-ionization rate from the effective photo-ionization rate leaves the diffuse photo-ionization rate.  The on-the-spot approximation produces a constant ratio of approximately 0.6 between diffuse and central source photons.  When ray tracing is used to follow the isotropic emission of recombination photons, their contribution peaks near a radius of 4 kpc on the flat part of the ionization profile and goes to zero at the center of the sphere.  These results are consistent with the findings of \citet{Raicevic_14}. 

\subsection{ Helium Recombination Radiation }
\label{sec.he_rec}

Few papers have focused on the effects of diffuse radiation in the
presence of hydrogen and helium \citep[although see][]{Cantalupo_11}.
In \rabacus, we allow for the transport of radiation due to
recombinations to the ground state for all ions.  This allows for
meaningful comparison to case A and case B approximations.  In
  future versions of the code we will include ionizing photons
  produced by other processes such as Balmer series recombinations,
  two photon processes, and secondary ionizations and heating from
  fast ($E > 100$ eV) electrons \citep{Shull85, Furlanetto10}.

\begin{figure}
\begin{center}
\includegraphics[width=0.47\textwidth]
{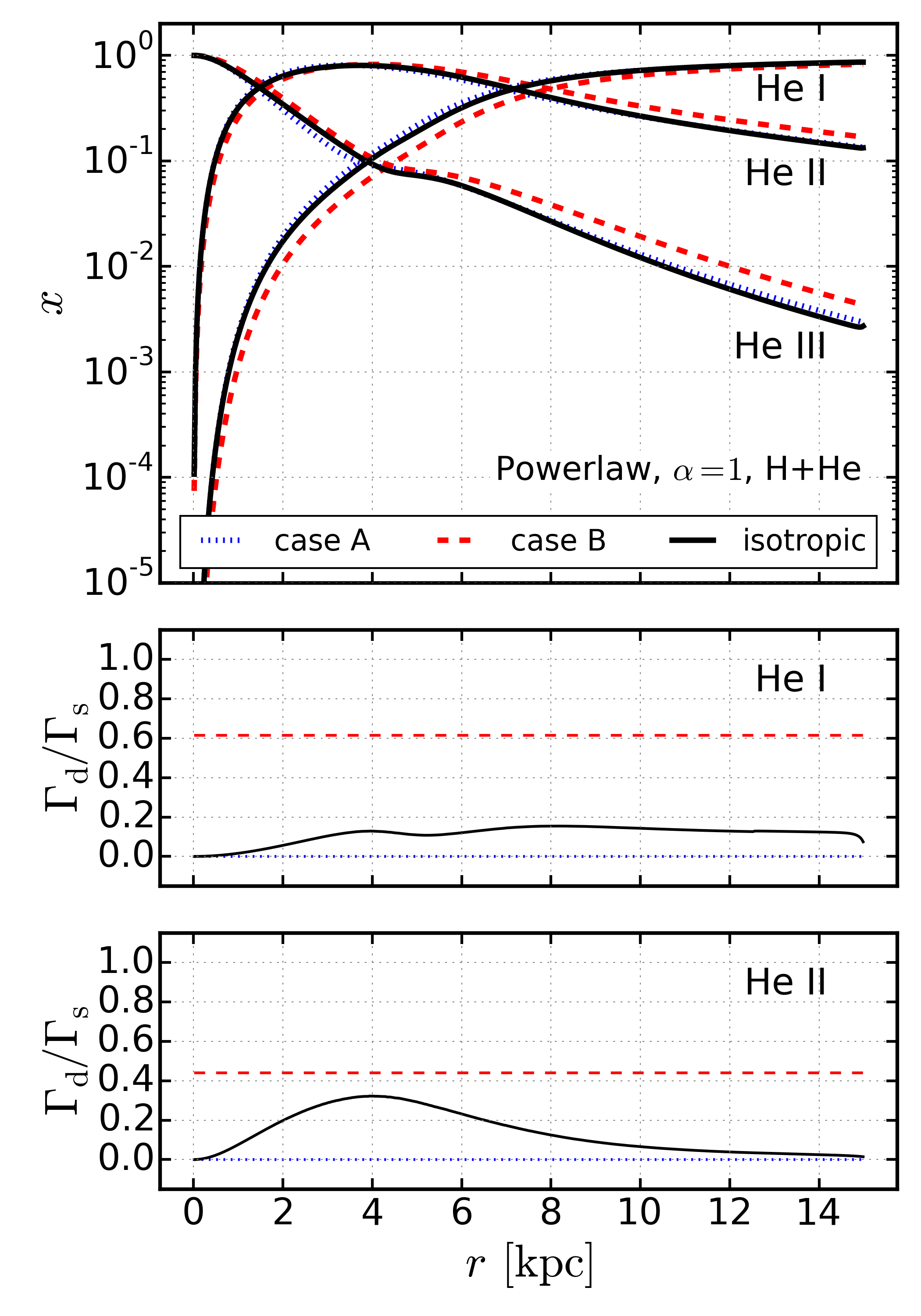}
\caption{ Results for a version of the test described in \protect
  \citet{Raicevic_14} extended to include helium.  In the upper panel
  we show the ionization fractions for all three helium ions resulting
  from using case A rates (ignoring recombination radiation), using
  case B rates (using the on-the-spot approximation) and from
  transporting recombination radiation isotropically.  In the lower
  panels we show the ratio of diffuse to central source
  photo-ionization rates.}
\label{fig.raicevic_He_poly}
\end{center}
\end{figure}

\begin{figure}
\begin{center}
\includegraphics[width=0.47\textwidth]
{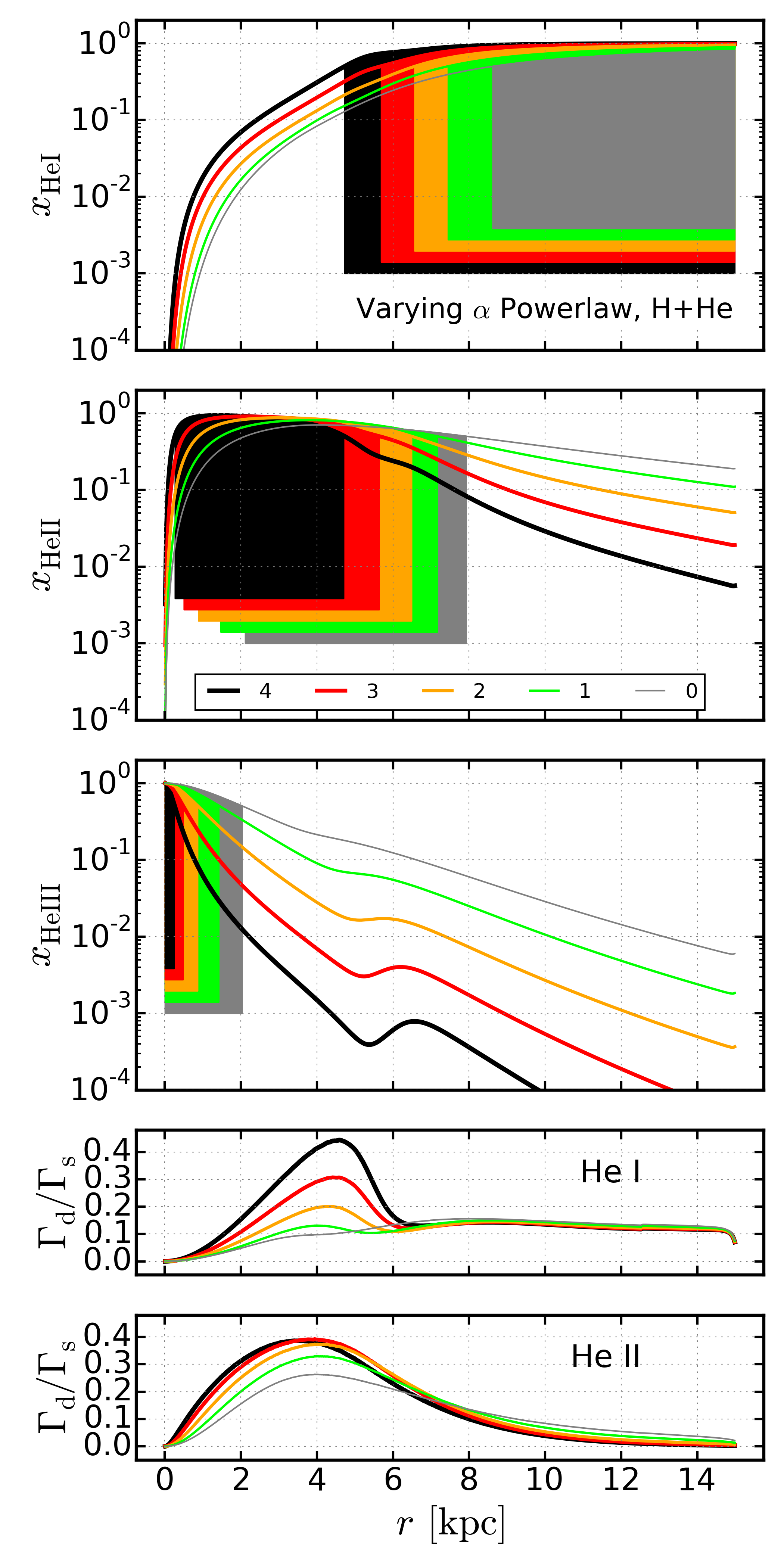}
\caption{ The effect of the powerlaw index, $\alpha$, on the helium
  ionization structure.  In the upper three panels we show the
  ionization fractions for neutral, singly ionized, and doubly ionized
  helium.  We show the results for five powerlaw spectra, $\Lnu
  \propto \nu^{-\alpha}$, with $\alpha$ ranging between 4 and 0 (in
  black, red, orange, lime, and grey respectively).  In each of these
  panels, the radii for which the ionization fraction is above 0.5 are
  indicated with a filled region.  In the lower panels we show the
  ratio of diffuse to central source photo-ionization rates.  As the
  spectrum becomes steeper, the contribution from recombination
  radiation becomes more important.  }
\label{fig.powerlaw_vary}
\end{center}
\end{figure}

To examine the effects of helium recombination photons we create a
point source with a photon luminosity of $L_{\rm n} = 10^{48}$ photons
per second and a powerlaw spectrum with $\Lnu \propto \nu^{-\alpha}$
between 1 and 60 Rydbergs.  Fig. \ref{fig.raicevic_He_poly} is
analogous to Fig. \ref{fig.raicevic_H_mono} but for helium ionization
profiles and photo-ionization rates.  For isotropic transport of
recombination photons we find that the diffuse photo-ionization rate
for neutral helium is never greater than 0.2 of the direct source rate
but peaks around 4 kpc for singly ionized helium as in the case of
hydrogen.  
While the case B approximation is more accurate for the outer
  radii in our previous tests involving hydrogen, the case A
  approximation is more accurate for helium.  This result is due to
  the fact that helium recombination photons are less effective at ionizing
  helium than hydrogen recombination photons are at ionizing
  hydrogen.  In fact, the opacity seen by helium recombination photons
  comes mostly from neutral hydrogen.  In addition, the higher energy
  and lower photo-ionization cross-section of helium recombination
  photons allow some to escape the system all together (see \S
  \ref{sec.ap_rec}).  However, we show in Fig. \ref{fig.powerlaw_vary}
  that these results are sensitive to the steepness of the central
  source spectrum.

To illustrate the previous point, we calculate the ionization
  structure in the sphere using five powerlaw spectra with spectral
  slopes $\alpha$ evenly spaced between 0 and 4 all normalized to emit
  $L_{\rm n} = 10^{48}$ photons per second.
  We present the results in Fig. \ref{fig.powerlaw_vary}.  In
  general, steeper spectral slopes produce a larger ratio of
  low-energy to high-energy photons leading to an increased
  singly-ionized helium fraction at $r < 4$ kpc.  This
  in turn leads to a higher rate of recombination for singly ionized
  helium and the increased importance of recombination radiation
  relative to central source photons for ionizing neutral helium in the
  inner region.  As the spectrum flattens, diffuse radiation
becomes decreasingly important for both neutral and singly ionized
helium, however the effect is stronger for neutral helium. 
  Furthermore, a flatter spectrum contains more high-energy photons
  than steeper spectra, which then doubly-ionize helium at a higher
  rate.  However, the optical depth of its recombination radiation is
  small (see \S\ref{sec.ap_rec}) and most of the diffuse radiation
  escapes from the system.

\subsection{ Isolating Physical Processes }

\begin{figure*}
\begin{center}
\includegraphics[width=0.94\textwidth]
{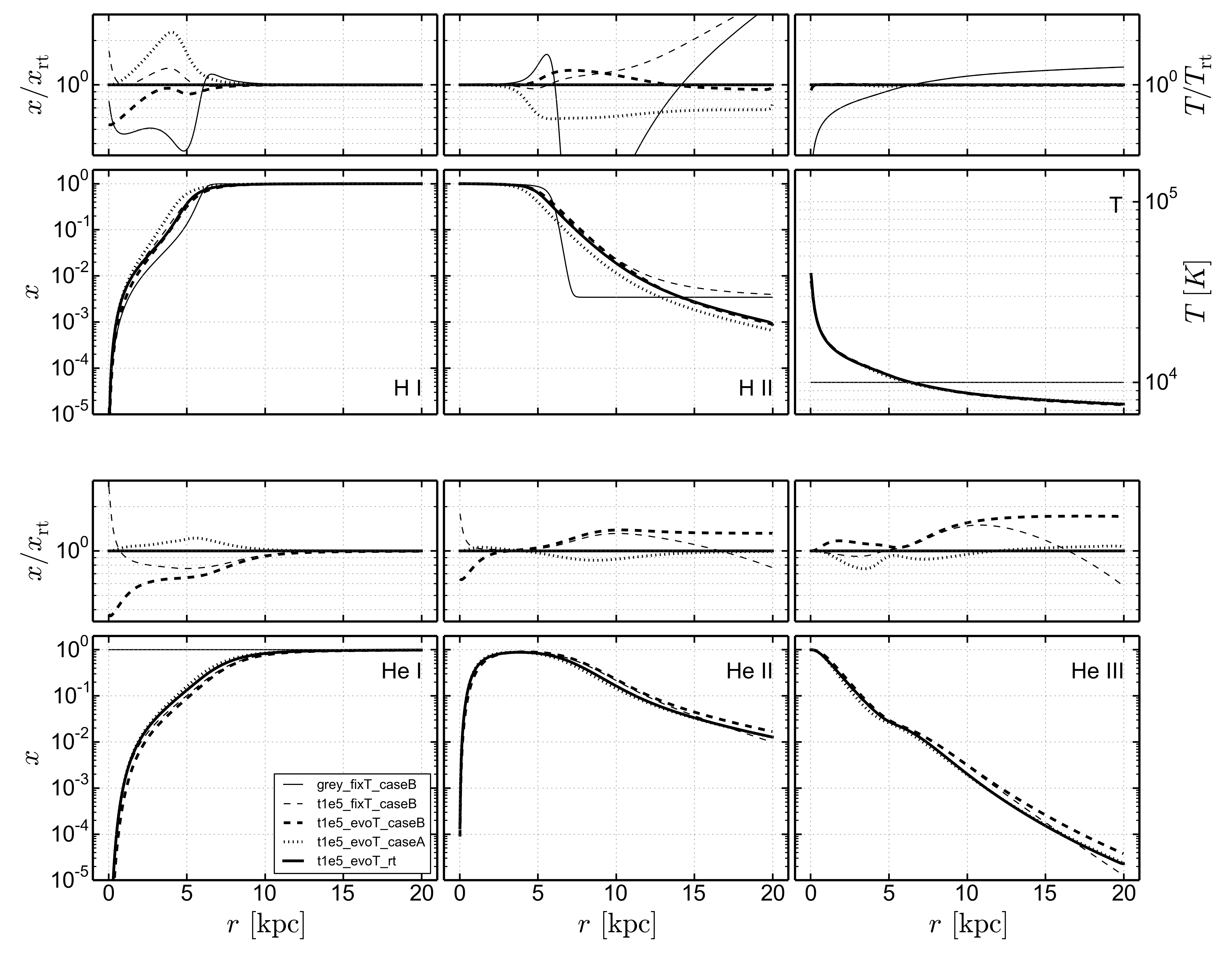}
\caption{ Ionization fraction and temperature profiles in five simple \protect{Str\"{o}mgren} sphere models.  Each panel shows the ionization fraction of one ion or the temperature.  The five models considered are\protect{:} 1) {\bf grey\_fixT\_caseB} (thin solid) in which the grey approximation to the $10^5$~K blackbody spectrum is used, the temperature is kept fixed, and recombination rates are case B, 2) {\bf t1e5\_fixT\_caseB} (thin dashed) in which the full $10^5$~K blackbody spectrum is used, the temperature is kept fixed, and recombination rates are case B, 3) {\bf t1e5\_evoT\_caseB} (thick dashed) in which the full $10^5$~K blackbody spectrum is used, equilibrium temperatures are calculated, and recombination rates are case B, 4) {\bf t1e5\_evoT\_A} (dotted) as in {\bf t1e5\_evoT\_B} but with case A recombination rates (i.e. recombination radiation is ignored) and 5) {\bf t1e5\_evoT\_rt} (thick solid) in which the full $10^5$~K blackbody spectrum is used, equilibrium temperatures are calculated, recombination rates are case A, and recombination radiation is transferred through the system as opposed to being absorbed on the spot.  Above each profile we show the ratio of each model's prediction to those of {\bf t1e5\_evoT\_rt} (i.e. dex of difference in the profile panels).  The scale in the top panels is a factor of three in either direction.  }
\label{fig.strom_phys_full_x}
\end{center}
\end{figure*}

As a final examination of the capabilities of the {\tt SphereStromgren} class we solve five models similar to those described in the slab geometry section (see \S \ref{sec.slab2_full}) to gain some insight into the effects of different physical processes.  The sphere we consider has a radius of 20 kpc and hydrogen and helium number densities of $\nH = 10^{-3}$~cm$^{-3}$ and $\nHe = 8.7 \times 10^{-5}$~cm$^{-3}$.  We consider a $10^5$~K blackbody source and its grey approximation (17.8 eV) both normalized to emit $5 \times 10^{48}$ photons per second.  The polychromatic models are prefixed with {\bf t1e5} as opposed to {\bf hm12} to indicate this change in spectrum. 

In Fig. \ref{fig.strom_phys_full_x} we show ionization and temperature profiles for all five models.  Above each profile panel we show the ratio of each model's predictions to those of {\bf t1e5\_evoT\_rt} (i.e. dex of difference in the bottom panel).  By comparing models {\bf grey\_fixT\_B} and {\bf t1e5\_fixT\_B} we can isolate the effect of using the grey approximation. The monochromatic model is characterized by a steep transition from ionized to neutral hydrogen at approximately 7 kpc which can be seen clearly in the $\xHII$ panel.  As in the case of the slab, using the polychromatic thermal spectrum smooths the hydrogen ionization profile.  We note that the two models reach the same $\xHII$ value at large radii where collisional ionizations dominate over photo-ionizations for hydrogen.  The harder photons that escape the ionized hydrogen region have a larger cross-section for absorption due to helium than from hydrogen. 

By comparing models {\bf t1e5\_fixT\_B} and {\bf t1e5\_evoT\_B} we can isolate the effect of self-consistently calculating the temperature structure. The equilibrium temperature is greater than 10$^4$~K in the region in which hydrogen is ionized, but falls below that temperature at larger radii.  For hydrogen, this produces lower neutral fractions at small radii and lower ionized fractions at large radii.  This is because the ionization state of hydrogen at radii $r>7$~kpc is determined by collisional ionizations.  For helium, the higher temperatures at small radii also reduce $\xHeI$, but helium ionizing photons still reach radii where the temperature is cooler than $10^4$~K and the reduced absorption of photons at small radii due to the higher temperatures leads to larger ionized fractions for both $\xHeII$ and $\xHeIII$ in model {\bf t1e5\_evoT\_B}.  

The larger case A recombination rate in model {\bf t1e5\_evoT\_A} produces more neutral species for both hydrogen and helium.  In model {\bf t1e5\_evoT\_rt} we calculate the transfer of recombination photons isotropically emitted from each layer.  As in the case of the slab, the rt and case A solutions agree in the optically thin center of the sphere.  Interestingly, the case B approximation is more accurate for hydrogen while the case A approximation is more accurate for helium.  Overall, the variations we have examined lead to differences of approximately a factor of 2 for the hydrogen ionization profiles and a factor of 50\% for the helium ionization profiles.

\section{Background Sphere Solver}

In this section we describe the {\tt SphereBgnd} class which models a
spherically symmetric gas distribution placed in an isotropic
background radiation field.  Even in the simplified case of
monochromatic radiation incident on a uniform density and temperature
sphere, there are no analytic solutions for the ionization profile.
In addition, the code we have used for comparison thus far, \cloudy,
does not support this geometry.

This geometry has been studied in the context of mini-halo evaporation
\citep[e.g.][]{Shapiro_04} and in the context of hydrogen absorption
systems \citep[e.g.][]{Zheng_02} however a direct comparison with either work
would not be fruitful.  In the first case, ionization profiles are not
presented.  In the second it would be difficult to know if
differences arise from the set of included physical processes, the
method of solution, or the implementation in code.  
Our goal for this final section is not a validation of the
{\tt SphereBgnd} class but rather a description of results in the
light of the other validated pieces of the \rabacus.  A complete
validation of this geometry must await a dedicated comparison
project.

In what follows we will describe a simple model for a
gaseous halo immersed in the cosmic UV background radiation.  Dark
matter halos have been shown to have universal density profiles
\citep{NFW_97} characterized by a scale radius $R_{\rm s}$.  To model
the gaseous component of the halo we create an NFW profile with a core
at $r=3R_{\rm s}/4$ as in \citet{Maller_04} and
\cite{Barnes_14}. Specifically, the virial radius $R_{\rm v}$ is
related to the virial mass $M_{\rm v}$ as, 
\begin{equation}
R_{\rm v} = 46.1 \, {\rm kpc} \, 
\left( \frac{\Delta_{\rm v} \Omega_{\rm m} h^2}{24.4} \right)^{-1/3}
\left( \frac{1+z}{3.3} \right)^{-1}
\left( \frac{M_{\rm v}}{10^{11} \Msun} \right)^{1/3} 
\end{equation}
where $\Delta_{\rm v}$ is the mean overdensity inside the virial
radius of the halo.  The ratio of the virial radius and the scale
radius is known as the concentration parameter $c = R_{\rm v} / R_{\rm
  s}$.  For our simple model we take the average relationship between
concentration and virial mass described in \citet{Maccio_07},
\begin{equation}
c = c_0 \left( \frac{M_{\rm v}}{10^{11} \Msun} \right)^{-0.109}
\left( \frac{1+z}{4} \right)^{-1} 
\end{equation}
with $c_0 = 3.5$. The gaseous hydrogen density profile is, 
\begin{equation} 
  \nH(r) = \frac{R^3_{\rm s} \rho_0 }{[r+(3/4) R_{\rm s}](r+R_{\rm s})^2}
\end{equation}
where $\rho_0$ is a constant set such that the integrated hydrogen gas
mass is equal to $M_{\rm H} = f_{\rm H} M_{\rm v}$ and $f_{\rm H} =
(1-Y_{\rm p}) \Omega_{\rm b} / \Omega_{\rm m}$.  The profile is then
fully specified by choosing a virial mass, a mean overdensity inside
the viral radius, and a redshift which we take to be $M_{\rm v} =
10^{10} \Msun$, $\Delta_{\rm v}=200$, and $z=0$ for our example.  This
results in a concentration parameter of 18. 

\begin{figure*}
\begin{center}
\includegraphics[width=0.94\textwidth]
{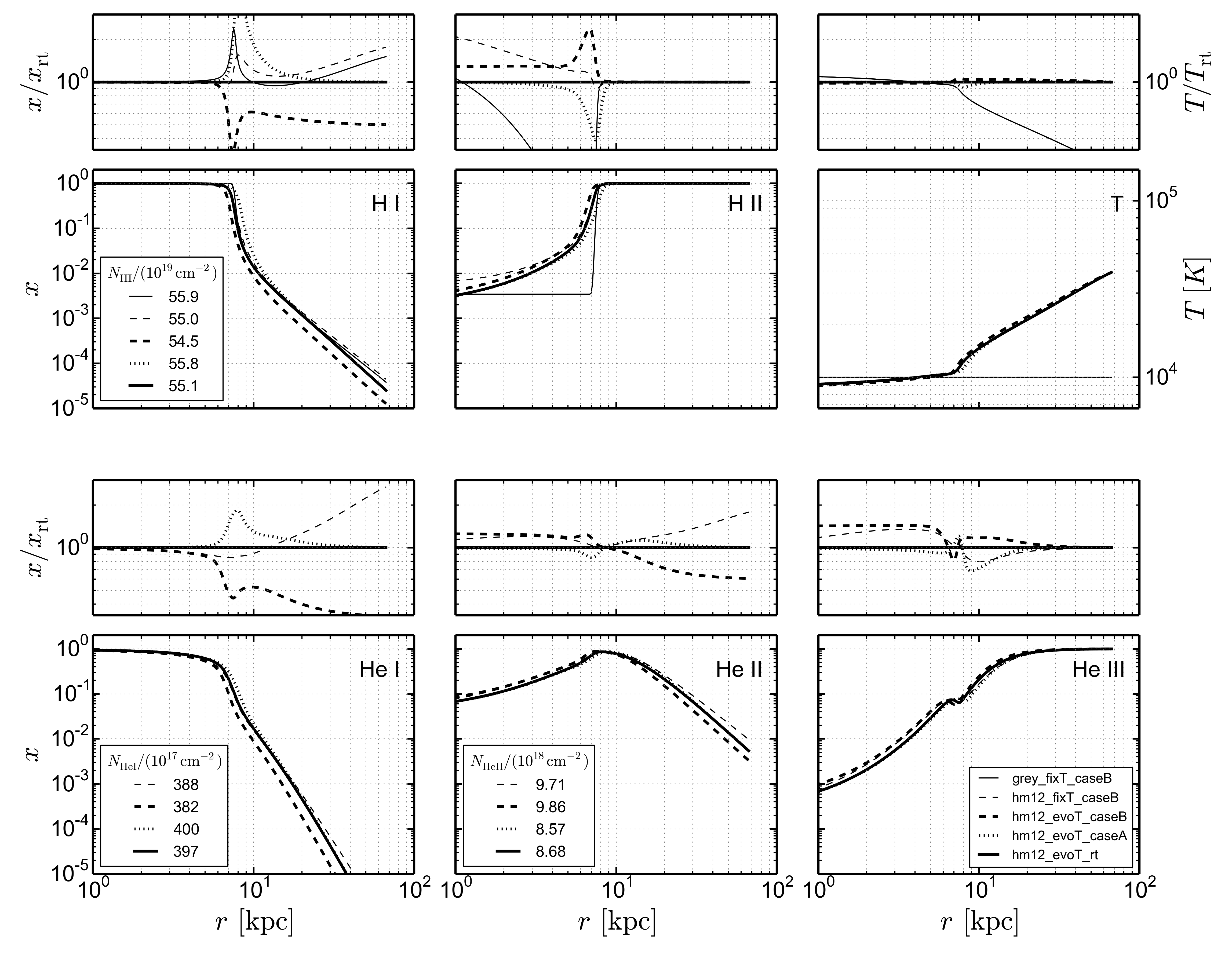}
\caption{ Ionization fraction and temperature profiles in five simple models of a gaseous halo in the UV background.  Each panel shows the ionization fraction of one ion or the temperature.  The five models considered are\protect{:} 1) {\bf grey\_fixT\_caseB} (thin solid) in which the grey approximation to the HM12 spectrum is used, the temperature is kept fixed, and recombination rates are case B, 2) {\bf hm12\_fixT\_caseB} (thin dashed) in which the full HM12 spectrum is used, the temperature is kept fixed, and recombination rates are case B, 3) {\bf hm12\_evoT\_caseB} (thick dashed) in which the full HM12 spectrum is used, equilibrium temperatures are calculated, and recombination rates are case B, 4) {\bf hm12\_evoT\_A} (dotted) as in {\bf hm12\_evoT\_B} but with case A recombination rates (i.e. recombination radiation is ignored), and 5) {\bf hm12\_evoT\_rt} (thick solid) in which the full HM12 spectrum is used, equilibrium temperatures are calculated, recombination rates are case A, and recombination radiation is transferred through the system as opposed to being absorbed on the spot.  Above each profile we show the ratio of each model's prediction to those of {\bf hm12\_evoT\_rt} (i.e. dex of difference in the profile panels). The scale in the top panels is a factor of three in either direction. }
\label{fig.sphere_bgnd_full_x}
\end{center}
\end{figure*}

The ray tracing tools necessary to calculate the transfer of the
isotropic background through the spherical geometry are the same we
described in the context of isotropic recombination radiation.  In
other words, any ray we need to trace for treating an isotropic
background can also, through the appropriate rotation, be made to be
perpendicular to the vertical axis. 
In Fig. \ref{fig.sphere_bgnd_full_x} we show ionization and temperature
profiles using the same five models as in Fig. \ref{fig.slab2_full_x}:
{\bf grey\_fixT\_B}, {\bf hm12\_fixT\_B}, {\bf hm12\_evoT\_B}, {\bf
  hm12\_evoT\_A}, and {\bf hm12\_evoT\_rt}.  In this case, the steep
density profile serves to lessen the differences between the models
relative to the constant density slab case.  This is partly due to the
fact that the chosen halo mass produces a neutral core for all
models.  Models are most sensitive to changes when the core is near
the optically thin to optically thick transition. 
Examining in detail the contribution to absorber abundances from
different halo masses is beyond the scope of this paper, but one of
the primary intended applications of \rabacus.

\section{Conclusion}

We have described \rabacus, a Python package for calculating the transfer of hydrogen ionizing radiation in simplified geometries relevant to astronomy and cosmology.  We presented example solutions for: 1) a semi-infinite slab gas distribution with isotropic radiation incident from both sides, 2) a spherically symmetric gas distribution with a point source at the center, and 3)  a spherically symmetric gas distribution in a uniform background.   We have also demonstrated the effects of temperature evolution, the inclusion of helium, and the treatment of recombination radiation on the ionization and temperature profiles in these geometries.  For slabs exposed to isotropic backgrounds at Lyman limit system densities, these processes can change hydrogen and helium column densities by an order of magnitude.  In the case of spherical geometries the magnitude of these effects is not as large but can still change ionization fractions by factors of 2.  The flexibility and speed of \rabacus\ lend itself to rapid prototyping of ideas and allows it to serve as a check for more complicated radiative transfer schemes.  It is made publicly available and is written in an interpreted language, Python, with a large user base.  Future improvements will include the addition of metal line cooling and non-equilibrium solvers.


\section*{Acknowledgments}

We thank an anonymous referee for an insightful review that helped
improve this manuscript.  JHW acknowledges support by NSF grants
AST-1211626 and AST-1333360.

\appendix

\section{Recombination Radiation}
\label{sec.ap_rec}

\begin{figure}
\begin{center}
\includegraphics[width=0.47\textwidth]
{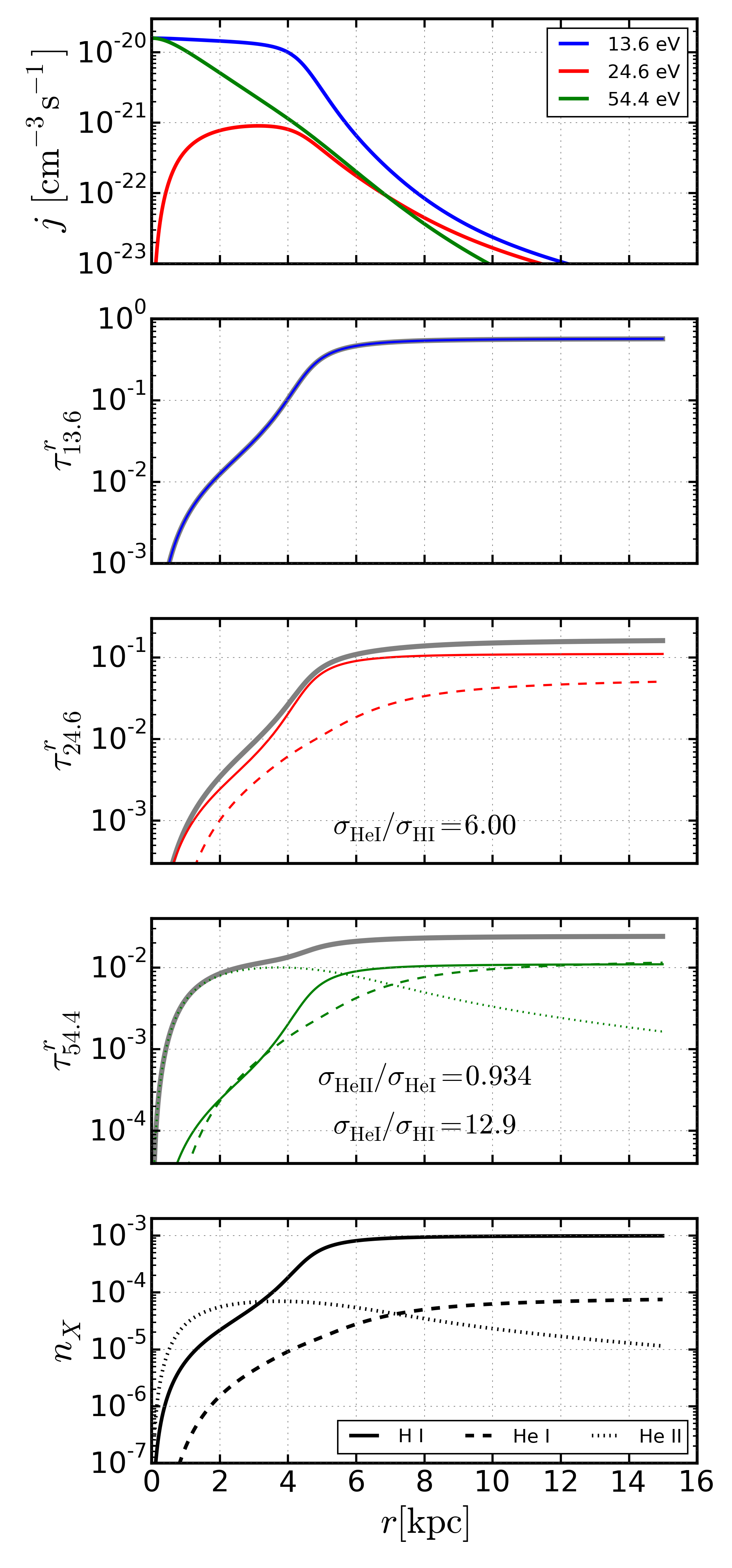}
\caption{ The opacity encountered by recombination radiation in the Stromgren sphere test of in \S \ref{sec.he_rec}. The top panel shows the emission coefficients as calculated in Eqs. \ref{eq.jcoefH} and \ref{eq.jcoefHe}.  The three middle panels show $\tau^{r}$ (see Eq. \ref{eq.tau_r}) as a function of radius at the ionization energies of hydrogen and helium. The total $\tau^{r}$ (solid grey line) at each radius is the sum of contributions from neutral hydrogen, neutral helium and singly ionized helium (solid, dashed, and dotted lines respectively).  The bottom panel shows the number density of each ion as a function of radius.  Hydrogen absorbs the majority of recombination radiation at 13.6 eV and 24.6 eV.  For recombination radiation at 54.4 eV, singly ionized helium absorbs the majority of photons at the inner radii while neutral hydrogen and helium absorb the majority at larger radii.  }
\label{fig.ref13}
\end{center}
\end{figure}

In this section we examine in more detail the Stromgren sphere test described in \S \ref{sec.he_rec}.  Our goal is to determine the total opacity encountered by recombination photons and the relative contribution to the total from different ions. We will make use of the quantity  $\tau^{r}$ which is simply the optical depth through each thin radial shell of thickness $dr$,  
\begin{align}
\label{eq.tau_r}
\tau^{\rm r}_{13.6} / dr &= 
\nHI \sigma_{\rm _{HI}}(E_{\rm _{HI}}) \nonumber \\
\tau^{\rm r}_{24.6} / dr &= 
\nHI \sigma_{\rm _{HI}}(E_{\rm _{HeI}}) + 
\nHeI \sigma_{\rm _{HeI}}(E_{\rm _{HeI}}) \nonumber \\
\tau^{\rm r}_{54.4} / dr &= 
\nHI \sigma_{\rm _{HI}}(E_{\rm _{HeII}}) + 
\nHeI \sigma_{\rm _{HeI}}(E_{\rm _{HeII}}) + 
\nHeII \sigma_{\rm _{HeII}}(E_{\rm_{HeII}})                     
\end{align}
In Fig. \ref{fig.ref13} we show the radial dependence of several quantities.  The top panel shows the emission coefficients as calculated in Eqs. \ref{eq.jcoefH} and \ref{eq.jcoefHe}.  The majority of recombination photons emitted at any radii come from hydrogen.  At inner radii, recombinations of doubly ionized helium produce the majority of helium recombination photons while at radii greater than 5 kpc both types of helium recombination produce a similar number of photons.  

\begin{figure}
\begin{center}
\includegraphics[width=0.47\textwidth]
{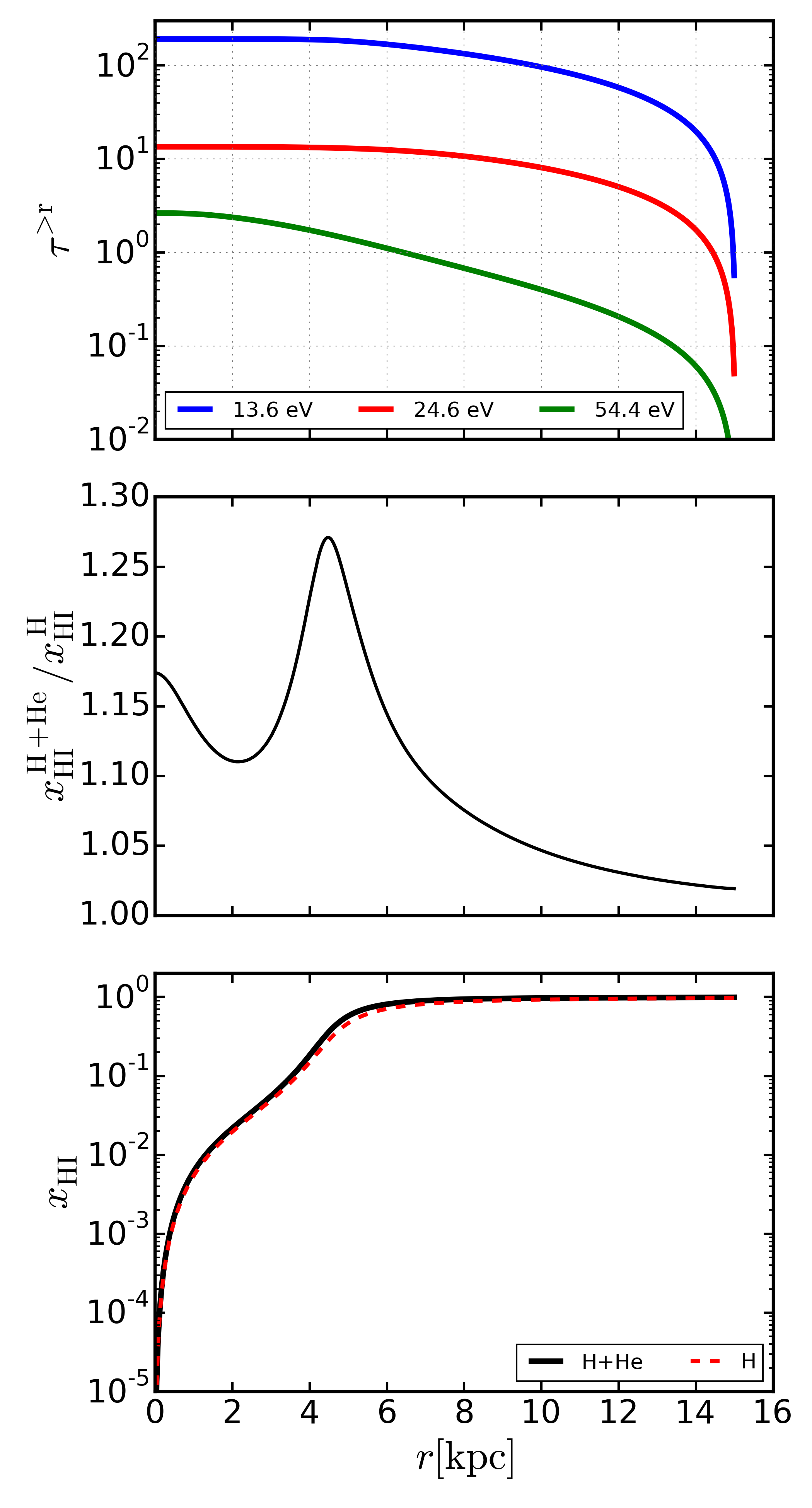}
\caption{ The effect of helium on the ionization profile of hydrogen in the Stromgren sphere test in \S \ref{sec.he_rec}.  In the top panel we show $\tau^{>r}$ (see Eq. \ref{eq.tau_gtr}) for the three ionization threshold energies of hydrogen and helium.  This can be used to understand how many recombination photons escape the system.  In the bottom two panels we show the neutral hydrogen ionization fraction profile with and without helium.  The inclusion of helium enhances $\xHI$ by as much as 25\%. }
\label{fig.ref13_2}
\end{center}
\end{figure}

In the middle three panels we show how the total opacity is partitioned amongst the absorbing species.  Photons at 13.6 eV are only absorbed by hydrogen and so the optical depth profile follows that of the $\nHI$ profile (see bottom panel in the same figure).  Photons at 24.6 eV are absorbed mostly by neutral hydrogen.  The cross-section for absorption by helium is six times larger than that of hydrogen at this energy, but the greater density of neutral hydrogen at all radii more than compensates for this. Photons at 54.4 eV can be absorbed by all three absorbing species.  The cross-sections at this energy for both helium species differ by less than ten percent, but singly ionized helium has a cross-section that is 13 times larger than that of hydrogen.  At small radii, singly ionized helium absorbs the majority of photons.  At larger radii, neutral hydrogen and helium absorb a similar amount of photons but the absorption by hydrogen is always equal to or greater than that by neutral helium.    
In Fig. \ref{fig.ref13_2} we examine changes to the hydrogen ionization profile caused by the inclusion of helium in \S \ref{sec.he_rec}.  There are two competing effects to consider.  Helium will absorb some of the source photons that would have been absorbed by hydrogen, however some fraction of the helium recombination photons will be absorbed by hydrogen.  In the bottom two panels of Fig. \ref{fig.ref13_2} we show that the first is larger and that the neutral hydrogen fraction increases by as much as twenty percent when helium is included.  Part of the reason that the helium photons do not make a larger contribution is because they escape the system.  This can be seen in the top panel in which we show $\tau^{>r}$.  
\begin{align}
\label{eq.tau_gtr}
\tau^{>r}_{13.6} &= \int_r^{\infty} \tau^{\rm r}_{13.6} \nonumber \\
\tau^{>r}_{24.6} &= \int_r^{\infty} \tau^{\rm r}_{24.6} \nonumber \\
\tau^{>r}_{54.4} &= \int_r^{\infty} \tau^{\rm r}_{54.4}
\end{align}
The optical depth to the surface of the sphere for photons at 54.4 eV is only larger than unity at radii smaller than 6 kpc.  While recombination photons at 24.6 eV do encounter significant optical depth, the abundance of these photons is always lower than those from hydrogen recombinations (see top panel of Fig. \ref{fig.ref13}).

\bibliographystyle{model2-names}
\bibliography{ms}







\end{document}

%% file: local-commands.tex
\newcommand{\rabacus}{\textsc{rabacus}}
\newcommand{\cloudy}{\textsc{cloudy}}

\newcommand{\Lnu}{L_{\nu} }
\newcommand{\Inu}{I_{\nu} }
\newcommand{\Nnu}{N_{\nu} }
\newcommand{\Jnu}{J_{\nu} }
\newcommand{\jnu}{j_{\nu} }
\newcommand{\unu}{u_{\nu}}
\newcommand{\taunu}{\tau_{\nu}}

\newcommand{\xHI}{x_{\rm _{HI}} }
\newcommand{\xHII}{x_{\rm _{HII}} }
\newcommand{\xe}{x_{\rm _{e}} }
\newcommand{\xHeI}{x_{\rm _{HeI}} }
\newcommand{\xHeII}{x_{\rm _{HeII}} }
\newcommand{\xHeIII}{x_{\rm _{HeIII}} }

\newcommand{\xHIce}{x_{\rm _{HI}}^{\rm ce} }
\newcommand{\xHIIce}{x_{\rm _{HII}}^{\rm ce} }
\newcommand{\xHeIce}{x_{\rm _{HeI}}^{\rm ce} }
\newcommand{\xHeIIce}{x_{\rm _{HeII}}^{\rm ce} }
\newcommand{\xHeIIIce}{x_{\rm _{HeIII}}^{\rm ce} }

\newcommand{\nH}{n_{\rm _{H}} }
\newcommand{\nHe}{n_{\rm _{He}} }
\newcommand{\nel}{n_{\rm e} }
\newcommand{\nelmax}{n_{\rm e}^{\rm max} }
\newcommand{\nelmin}{n_{\rm e}^{\rm min} }

\newcommand{\nHI}{n_{\rm _{HI}} }
\newcommand{\nHII}{n_{\rm _{HII}} }
\newcommand{\nHeI}{n_{\rm _{HeI}} }
\newcommand{\nHeII}{n_{\rm _{HeII}} }
\newcommand{\nHeIII}{n_{\rm _{HeIII}} }

\newcommand{\GHI}{\Gamma_{\rm _{HI}} }
\newcommand{\GHeI}{\Gamma_{\rm _{HeI}} }
\newcommand{\GHeII}{\Gamma_{\rm _{HeII}} }

\newcommand{\epHI}{\epsilon_{\rm _{HI}} }
\newcommand{\epHeI}{\epsilon_{\rm _{HeI}} }
\newcommand{\epHeII}{\epsilon_{\rm _{HeII}} }

\newcommand{\sigmaHI}{\sigma_{\rm _{HI}} }
\newcommand{\sigmaHeI}{\sigma_{\rm _{HeI}} }
\newcommand{\sigmaHeII}{\sigma_{\rm _{HeII}} }

\newcommand{\dtau}[2]{\Delta \tau_{ #1 , #2 } }

\newcommand{\NHI}{N_{\rm _{HI}} }
\newcommand{\NHeI}{N_{\rm _{HeI}} }
\newcommand{\NHeII}{N_{\rm _{HeII}} }

\newcommand{\nuHIth}{\nu_{\rm _{HI}}^{\rm th} }
\newcommand{\nuHeIth}{\nu_{\rm _{HeI}}^{\rm th} }
\newcommand{\nuHeIIth}{\nu_{\rm _{HeII}}^{\rm th} }

\newcommand{\CHI}{{\rm C}_{\rm _{HI}} }
\newcommand{\CHeI}{{\rm C}_{\rm _{HeI}} }
\newcommand{\CHeII}{{\rm C}_{\rm _{HeII}} }

\newcommand{\RHII}{{\rm R}_{\rm _{HII}} }
\newcommand{\RHeII}{{\rm R}_{\rm _{HeII}} }
\newcommand{\RHeIII}{{\rm R}_{\rm _{HeIII}} }

\newcommand{\mIHI}{\mathcal{I}_{\rm _{HI}} }
\newcommand{\mIHeI}{\mathcal{I}_{\rm _{HeI}} }
\newcommand{\mIHeII}{\mathcal{I}_{\rm _{HeII}} }

\newcommand{\mRHII}{\mathcal{R}_{\rm _{HII}} }
\newcommand{\mRHeII}{\mathcal{R}_{\rm _{HeII}} }
\newcommand{\mRHeIII}{\mathcal{R}_{\rm _{HeIII}} }

\newcommand{\myvec}[1]{\mathbf{#1}}

\newcommand{\Msun}{{\rm M}_{\odot}}

\newcommand{\cubecm}{\ifmmode{{\rm cm^{-3}}}\else{cm$^{-3}$}\fi}

\newcommand\abs[1]{\left|#1\right|}

\DeclareFontFamily{\encodingdefault}{\ttdefault}{\hyphenchar\font=`\-}